\documentstyle[12pt,epsf]{article}

\newcommand{\gtrsim}{ \mathop{}_{\textstyle \sim}^{\textstyle >} }
\newcommand{\lesssim}{ \mathop{}_{\textstyle \sim}^{\textstyle <} }

\begin{document}
\begin{titlepage}
\begin{center}

January 7, 1997
\hfill    LBNL-39753\\
\hfill UCB-PTH-96/63  \\
\hfill hep-ph/9701244\\

\vskip .25in

{\Large \bf Cosmology of Supersymmetric Models\\
with Low-energy Gauge Mediation}\footnote{This 
work was supported in part by the Director, Office of 
Energy Research, Office of High Energy and Nuclear Physics, Division 
of 
High Energy Physics of the U.S. Department of Energy under Contract 
DE-AC03-76SF00098 and in part by the National Science Foundation 
under 
grant PHY-95-14797.  HM was also supported by the Alfred P. Sloan 
Foundation.  AdG was supported by CNPq (Brazil).}

\vskip 0.3in

Andr\'e de Gouv\^ea$^1$, Takeo Moroi$^2$ and Hitoshi Murayama$^{1,2}$

\vskip 0.1in

{{}$^1$ \em Department of Physics\\
     University of California, Berkeley, California 94720}

\vskip 0.1in

{{}$^2$ \em Theoretical Physics Group\\
     Earnest Orlando Lawrence Berkeley National Laboratory\\
     University of California, Berkeley, California 94720}

\end{center}

\vskip .1in

\begin{abstract}

We study the cosmology of supersymmetric models in which the supersymmetry
breaking effects are mediated by gauge interactions at about the
$10^{5}$~GeV scale.  We first point out that the gravitino is likely
to overclose the Universe in this class of models.  This requires an
entropy production, which prefers a baryogenesis mechanism at a
relatively low temperature. The Affleck-Dine mechanism for
baryogenesis is one of the possibilities to generate enough baryon
asymmetry, but the analysis is non-trivial since the shape of the
potential for the flat direction differs substantially
from the conventional hidden sector case. To see this, we first perform a
2-loop calculation to determine the shape of the potential. By
combining the potential with the supergravity contribution, we then
find that the Affleck-Dine baryogenesis works efficiently to generate
sufficient baryon asymmetry.  On the other hand, we also point out
that string moduli fields, if present, are stable and their
coherent oscillations overclose the Universe by more than 15 orders of
magnitude.  One needs a very late inflationary period with an $e$-folding of $N
\gtrsim 5$ and an energy density of $\lesssim (10^{7}$~GeV$)^4$. A
thermal inflation is enough for this purpose.  Fortunately, the
Affleck--Dine baryogenesis is so efficient that enough baryon
asymmetry can survive the late inflation.

\end{abstract}

\end{titlepage}

\renewcommand{\thepage}{\roman{page}}
\setcounter{page}{2}
\mbox{ }

\vskip 1in

\begin{center}
{\bf Disclaimer}
\end{center}
\vskip .2in
\begin{scriptsize}
\begin{quotation}
This document was prepared as an account of work sponsored by the
United States Government. While this document is believed to contain
correct information, neither the United States Government nor any
agency thereof, nor The Regents of the University of California, nor
any of their employees, makes any warranty, express or implied, or
assumes any legal liability or responsibility for the accuracy,
completeness, or usefulness of any information, apparatus, product, or
process disclosed, or represents that its use would not infringe
privately owned rights.  Reference herein to any specific commercial
products process, or service by its trade name, trademark,
manufacturer, or otherwise, does not necessarily constitute or imply
its endorsement, recommendation, or favoring by the United States
Government or any agency thereof, or The Regents of the University of
California.  The views and opinions of authors expressed herein do not
necessarily state or reflect those of the United States Government or
any agency thereof, or The Regents of the University of California.
\end{quotation}
\end{scriptsize}
\vskip 2in

\begin{center}
\begin{small}
{\it Lawrence Berkeley Laboratory is an equal opportunity employer.}
\end{small}
\end{center}
\newpage

\renewcommand{\thepage}{\arabic{page}}
\setcounter{page}{1}
\setcounter{footnote}{0}

\section{Introduction}
\label{sec:intro}
\setcounter{equation}{0}
\setcounter{footnote}{0}

The promise of low-energy supersymmetry (SUSY) is to stabilize the
hierarchy between the weak scale and a higher scale of new physics, {\it
e.g.}\/ the Planck scale (see, {\it e.g.}\/, \cite{INS}).  Supersymmetry,
however, has to be spontaneously broken because we do not see
degenerate pairs of particles and their superpartners.  Moreover,
there are stringent phenomenological constraints on the spectrum of
superparticles such as the degeneracy among squarks or sleptons at the
percent level (see, {\it e.g.}\/, \cite{Masiero}).  Therefore,
constructing viable mechanisms of supersymmetry breaking has been
regarded as one of the most important issues in supersymmetry model
building.

Recently, the idea of generating supersymmetry-breaking masses via gauge
interactions has attracted interests (Low-Energy Gauge Mediation, or
LEGM) \cite{DN1,DN2,DN3}.  In this scheme, the supersymmetry breaking
effects appear in the supersymmetric Standard Model in the following
manner.  There is a sector which breaks supersymmetry dynamically at
around a $10^{7}$~GeV scale, and it generates supersymmetry breaking
effects in the so-called messenger sector at around $10^{5}$~GeV, which
further induce supersymmetry breaking masses of order $10^{3}$~GeV in
the supersymmetric Standard Model via ordinary gauge interactions.
This mechanism guarantees the required degeneracy among squarks,
sleptons at a high degree, and also generates the masses for scalars
and gauginos at comparable magnitudes as desired phenomenologically.
There have been active studies on the phenomenology of such models
\cite{CM,collider}.  On the other hand, there has been little discussion
on cosmological consequences of this mechanism, except issues
concerning stable particles in the messenger sector \cite{dark}.
Since the scheme completely differs from the conventional hidden sector
scenario at high energies, early cosmology is expected to differ
substantially as well.

There are (at least) two ingredients in the LEGM models which may lead
to a cosmology different from the hidden sector case.  The first is a
very small gravitino mass.  Since supersymmetry is broken at around
$10^{7}$~GeV, compared to around $10^{10}$~GeV in the hidden sector
scenario, the gravitino mass is much lighter: $m_{3/2} \sim 100$~keV
compared to $100$~GeV. The second is that the supersymmetry breaking
effects ``shut off'' at high energies.\footnote{A similar effect was
discussed in \cite{Nemeschansky} in the context of the sliding singlet
mechanism.} In particular, the flat directions in the supersymmetric
Standard Model have very different potentials at large field
amplitudes.

In this paper, we study the implications of the LEGM models to cosmology.
In Section~\ref{sec:MMY}, we first discuss the cosmological constraints
on light gravitinos mainly based on the analysis by two of the authors
(TM and HM) and Yamaguchi \cite{MMY}.  Then, in
Section~\ref{sec:gravitino_LEGM}, we pay particular attention to the
estimate of the gravitino mass in the LEGM models, and argue that it is
highly unlikely to be lighter than $2h^2{\rm keV}$ as required by
cosmology.  This point implies that there must be a substantial
entropy production, which casts a concern on the baryon
asymmetry.  Therefore, we turn our attention to a possible mechanism
of baryogenesis at a relatively low temperature, using the idea by
Affleck and Dine \cite{AD}. The important point in the LEGM models is that
the SUSY breaking effects due to the messenger interaction ``shut
off'' at high energies. Therefore, in Section~\ref{sec:flat-dir}, we
performed a 2-loop calculation to determine the shape of the potential
for the flat direction.  In Section~\ref{sec:ADinLEGM}, we estimate the
possible value of the baryon-to-entropy ratio which can be induced by
the Affleck-Dine baryogenesis.  For a sufficiently large amplitude of
the flat direction, the potential is dominated by the supergravity
contribution rather than the LEGM contribution, and we will see that
the Affleck-Dine baryogenesis works well enough to explain the present
value of the baryon asymmetry.  Furthermore, in Section~\ref{sec:moduli},
we point out that the string moduli, if present within the LEGM models,
cause a serious problem because they are stable and their coherent
oscillations grossly overclose the Universe.  However we also point
out a possible solution to the problem.  Since the Affleck--Dine
baryogenesis is so efficient, the baryon asymmetry can survive the
enormous entropy production required to dilute the moduli fields,
possibly by thermal inflation \cite{LS}. Finally, in
Section~\ref{sec:conclusion}, we summarize our conclusions.

\section{Cosmology of a Light Gravitino}
\label{sec:MMY}
\setcounter{equation}{0}
\setcounter{footnote}{0}

In this section, we briefly review the cosmology with a light stable
gravitino \cite{MMY}.\footnote {In this paper, we assume the absolute
conservation of $R$-parity.}  If a stable gravitino is thermalized in
the early Universe, and if it is not diluted by some mechanism (such as
a late inflation and/or a substantial entropy production), its mass
density may exceed the closure limit: $\Omega_{3/2}<1$. Since the number
density of the gravitino is fixed once the gravitinos are thermalized, the
above argument sets an upper bound on the gravitino mass~\cite{PP}:
 \begin{eqnarray}
m_{3/2}\lesssim 2 h^2{\rm keV}
{\rm ~:~without~dilution},
\label{1kev}
 \end{eqnarray} 
where $h$ is the Hubble constant in units of 100km/sec/Mpc.  In other
words, if the gravitino mass is heavier than $2h^2$ keV, we need
some mechanism to dilute the gravitino in order not to overclose the
Universe. Since the gravitinos are produced more at a higher
temperature, we obtain an upper bound on the 
maximal temperature, $T_{\rm max}$, from which the ordinary radiation dominated
Universe starts. For example, in the inflationary Universe, $T_{\rm
max}$ corresponds to the so-called reheating temperature $T_{\rm RH}$ which is
typically higher than $T_{\rm RH}\gtrsim 10^8$ GeV, if there is no significant
entropy production after reheating.  If $T_{\rm
max}$ turns out to be less than $10^8$~GeV or so, we judge that one
needs a substantial entropy production below $T_{\rm RH}$.
It is worth to recall that the recent measurements prefer 
$h \sim 0.7$ and hence the upper bound is about 1~keV.  

The crucial point about the light gravitino is that the interaction of
the (longitudinal component of) gravitino becomes stronger as the
gravitino mass gets lighter. This is because the longitudinal
component of the gravitino behaves like the goldstino, whose
interaction is proportional to $\langle F\rangle^{-1}\sim
(m_{3/2}M_*)^{-1}$, where $M_{*} = 2.4\times 10^{18}$~GeV is the 
reduced Planck scale.  For the light gravitino, the interaction of the
longitudinal component of the gravitino ($\sim$ the goldstino) $\psi$
to the chiral multiplet ($\phi$, $\chi$) and to the gauge multiplet
($A_\mu$, $\lambda$) is given by \cite{Fayet}
 \begin{eqnarray}
{\cal L} = \frac{im_\lambda}{8\sqrt{6}m_{3/2}M_*}
\bar{\psi}[\gamma_\mu\gamma_\nu ]\lambda F^{\mu\nu}
+ \frac{m_\chi^2-m_\phi^2}{\sqrt{3}m_{3/2}M_*}
(\bar{\psi}\chi_{\rm L})\phi^* + {\rm h.c.},
\label{L_int}
 \end{eqnarray}
 where $m_\phi$, $m_\chi$, and $m_\lambda$ represent the masses of
$\phi$, $\chi$, and $\lambda$.\footnote
{Here, $\psi$ represents the spin $\frac{1}{2}$ field, though the
gravitino has spin $\frac{3}{2}$. In the high energy limit, 
$\psi$ is related to the
longitudinal (helicity $\pm 1/2$) component of the gravitino,
$\psi^\mu_{1/2}$, as $\psi^\mu_{1/2}\sim\sqrt{2/3}\partial^\mu\psi
/m_{3/2}$.} 
As indicated in Eq. (\ref{L_int}), the interaction of $\psi$ becomes
stronger as the gravitino mass gets smaller.

In the thermal bath, two types of the processes may contribute to
overproduce the gravitino: one is the decay of the sparticle
$\tilde{X}$ into its superpartner $X$ and the gravitino,
$\tilde{X}\rightarrow\psi +X$, and the other is the scattering
processes, $x+y\rightarrow\psi +z$, where $x,y,z$ are relevant
(s)particles. The decay process is significant especially for the case 
$m_{3/2}\lesssim$ 100keV. The partial decay width
of a sparticle $\tilde{X}$ into the gravitino is estimated as
 \begin{eqnarray}
\Gamma(\tilde{X}\rightarrow\psi +X)\sim
\frac{1}{48\pi}\frac{m_{\tilde{X}}^5}{m_{3/2}^2M_*^2},
 \label{Gamma_3/2}
 \end{eqnarray}
with $m_{\tilde{X}}$ being the mass of $\tilde{X}$, and it becomes
large as the gravitino mass gets small.  This decay process produces
gravitinos as $\dot{n}_{3/2} + 3 H n_{3/2} = \Gamma(\tilde{X}
\rightarrow \psi +X) n_{\tilde{X}}$ where $H$ is the expansion rate of
the Universe at the given time.\footnote{If the gravitino number
density is large, there is also a damping term because of the detailed
balance, $\mbox{(r.h.s.)} = \Gamma(\tilde{X} \rightarrow \psi +
X)n_{\tilde{X}}(1 - n_{3/2}/n_{3/2}^{\rm eq})$, where $n_{3/2}^{\rm
eq}$ is the thermal equilibrium value of $n_{3/2}$.}
Here and below, $n_{3/2}$ is the number density of gravitinos in the
Universe at a given time. 
If the gravitino mass is in the range $2h^2{\rm keV}\lesssim
m_{3/2}\lesssim 100{\rm keV}$, the decay rate becomes so large that
the decay process overproduces the gravitino once the sparticles are
thermalized \cite{MMY}. 
Thus, if the gravitino mass is in this range, $T_{\rm
max}$ should be lower than about $m_{\tilde{X}}\sim$ 100 GeV -- 1 TeV
depending on the mass spectrum of superparticles, or the
Universe is overclosed.\footnote
{A similar argument can be applied to the decay process of the
particles in the messenger sector or the SUSY breaking sector. In that
case, the decay rate becomes much larger since the  parent particle
is much heavier. Thus, if the particles in those sectors are
thermalized, the lower bound on the gravitino mass becomes more
stringent than $\sim 100{\rm keV}$.}

If the gravitino mass is heavier than $O(100{\rm keV})$, the
decay process becomes unimportant and the most important production
mechanisms of gravitinos are scattering processes. In this case, the Boltzmann
equation for the number density of the gravitino, $n_{3/2}$, is given
by
 \begin{eqnarray}
\dot{n}_{3/2}+3Hn_{3/2}=\Sigma_{\rm tot}n_{\rm rad}^2,
\label{boltzmann}
 \end{eqnarray}
where $H$ is the expansion rate of the Universe, $\Sigma_{\rm tot}$ is
the thermally averaged total cross section, and $n_{\rm rad}=(\zeta
(3)/\pi^2)T^3$. At high energies the first term in
Eq.~(\ref{L_int}) becomes more significant than the second one, and
hence $\Sigma_{\rm tot}$ is as large as $O(g_3^2m_{\rm
G3}^{2}/m_{3/2}^2M_*^2)$. After a detailed calculation, we
obtain~\cite{MMY}
 \begin{eqnarray}
\Sigma_{\rm tot}\sim
5.9 \frac{g_3^2m_{\rm G3}^{2}}{m_{3/2}^2M_*^2},
\label{Sigma_tot}
 \end{eqnarray}
where $g_3$ and $m_{\rm G3}$ are the gauge coupling constant and the
gaugino mass for SU(3)$_{\rm C}$.  Solving Eq.~(\ref{boltzmann}), and
taking account of the effect of the dilution factor,
$g_*(T_{\rm max})/g_*(T)$ (where $g_*(T)$ is the number of the
relativistic degrees of freedom in the thermal bath with temperature
$T$), the number density of the gravitino is given by
 \begin{eqnarray}
\frac{n_{3/2}(T)}{n_{\rm rad}}
&=&
\frac{g_*(T)}{g_*(T_{\rm max})}
\left. \frac{\Sigma_{\rm tot}n_{\rm rad}}{H}\right|_{T=T_{\rm max}}
\nonumber \\ &\sim&
3\times 10^{-2}
\frac{g_*(T)}{g_*(T_{\rm max})}
\left(\frac{m_{3/2}}{100{\rm keV}}\right)^{-2}
\left(\frac{m_{\rm G3}}{1{\rm TeV}}\right)^2
\left(\frac{T_{\rm max}}{10{\rm TeV}}\right).
 \end{eqnarray}
Using $g_*(T\lesssim 1{\rm MeV})\sim 3.9$, and 
$g_*(T_{\rm max})\sim 200$, we obtain
 \begin{eqnarray}
\Omega_{3/2}= \frac{m_{3/2}n_{3/2}}{\rho_{\rm c}}\sim 
1\times h^{-2}\left(\frac{m_{3/2}}{100 {\rm keV}}\right)^{-1}
\left(\frac{m_{\rm G3}}{1{\rm TeV}}\right)^2
\left(\frac{T_{\rm max}}{10{\rm TeV}}\right),
 \end{eqnarray}
and the condition $\Omega_{3/2}\leq 1$ sets an upper bound on 
$T_{\rm max}$. In summary, the
upper bound on $T_{\rm max}$ is given by~\cite{MMY}
 \begin{eqnarray}
T_{\rm max}\lesssim
\left\{
\begin{array}{ll}
{\rm 100 GeV~-~1TeV} &
{:~2h^2	{\rm keV}\lesssim m_{3/2}\lesssim 100 {\rm keV}}\\
{10{\rm TeV}\times 
h^2
\left(\frac{m_{3/2}}{100{\rm keV}}\right)
\left(\frac{m_{\rm G3}}{1{\rm TeV}}\right)^{-2}} &
{:~m_{3/2}\gtrsim 100 {\rm keV}}
\end{array}
\right. .
\label{Tmax}
 \end{eqnarray}

\begin{figure}
\centerline{\epsfxsize=0.8\textwidth \epsfbox{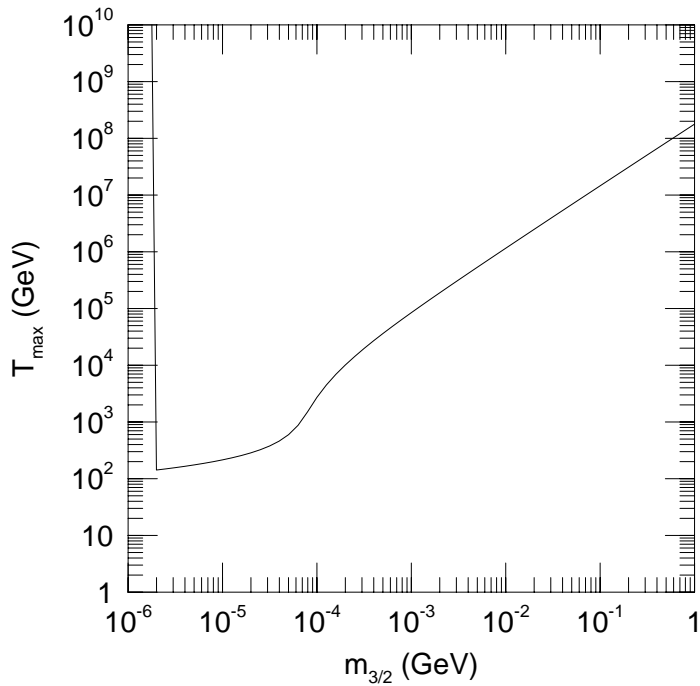}}
\caption{The upper bound on $T_{\rm max}$ as a function of the 
gravitino mass from the requirement that the relic stable gravitinos 
do not overclose the Universe.  We take the Hubble parameter to be 
$H_0 = 100~\mbox{Mpc}/\mbox{km}/\mbox{sec}$.  There is no constraint 
below $m_{3/2} = 2$~keV, which is represented by the vertical line.  
For smaller $H_0$, the constraints become more stringent.  The upper 
bound on $T_{\rm max}$ shifts towards smaller $T_{\rm max}$ as 
$(H_0)^2$.  The vertical line moves towards smaller $m_{3/2}$ also as 
$(H_0)^2$.  Note that the current data prefer $H_{0} \sim 
70~\mbox{Mpc}/\mbox{km}/\mbox{sec}$.  }
\label{fig:tr_max}
\end{figure}

The above constraints are summarized in the Fig.~\ref{fig:tr_max}.  As
one can see, the upper bound on $T_{\rm max}$ is much lower than the
usual reheating temperature after ordinary inflation, $T_{\rm RH}
\gtrsim 10^8$.  To reduce the number density of the gravitino, therefore,
a large entropy production is required.

\section{Light Gravitino in the LEGM Models}
\label{sec:gravitino_LEGM}
\setcounter{equation}{0}
\setcounter{footnote}{0}

We discussed cosmological constraints on a light stable gravitino in
the previous section, and showed that one needs to dilute gravitinos
produced in the early Universe somehow if $m_{3/2} \gtrsim
2h^{2}~\mbox{keV}$.  In this section we estimate the gravitino mass in
the LEGM models carefully and find it unlikely to be below $2h^2$ keV.

In the scheme of the LEGM models, there are three sectors: the dynamical SUSY
breaking (DSB) sector which originally breaks SUSY, the ordinary
sector which consists of the particles in the minimal SUSY standard model
(MSSM), and the messenger sector which mediates the SUSY breaking from
the DSB sector into the ordinary sector.  The scales for these sectors
have a large hierarchy, since they are related by loop factors:
$\Lambda_{\rm DSB}\gg\Lambda_{\rm mess}\gg M_{\rm SUSY}$, where
$\Lambda_{\rm DSB}$, $\Lambda_{\rm mess}$ and $M_{\rm SUSY}$
represent the scales for the dynamical SUSY breaking sector, messenger
sector, and the ordinary sector ($\sim$ electroweak scale), respectively.

In the LEGM models a gauge interaction, which becomes strong at the scale
$\Lambda_{\rm DSB}$, induces a non-perturbative superpotential.  
Due to non-perturbative effects, $F$-components of chiral
multiplets in the SUSY breaking sector acquire non-vanishing VEVs, $\langle
F_0\rangle\sim \Lambda_{\rm DSB}^2$, and SUSY is dynamically broken.
Assuming a vanishing cosmological constant, the gravitino mass in
this model is given by
 \begin{eqnarray}
m_{3/2}=\frac{\langle F_0\rangle}{\sqrt{3} M_*}
\sim\frac{\Lambda_{\rm DSB}^2}{M_*}.
 \end{eqnarray}

In the next stage, the SUSY breaking is fed down to the messenger
sector by integrating out the U(1)$_{\rm mess}$ interaction.  The
messenger sector contains a gauge singlet $S$, whose $A$- and
$F$-component $F_S$ acquire VEVs after minimizing the
potential.\footnote{One actually needs a substantially more complicated
messenger sector than the original ones \cite{DN1,DN2,DN3} in order to
avoid a run-away global minimum \cite{ACHM,DDR}.  Such details are,
however, beyond the scope of this paper.} The scale of these VEVs are
related to the original SUSY breaking scales $\Lambda_{\rm DSB}$ as
$\langle S\rangle\sim\langle F_S\rangle^{1/2}\sim O(g_{\rm
mess}^2\Lambda_{\rm DSB}/16\pi^2)$.  The ratio of $\langle F_S\rangle$
to $\langle S\rangle$ determines the masses of the sparticles in the
ordinary sector.  By counting the loop factors, we obtain
 \begin{eqnarray}
\Lambda_{\rm mess} \equiv
\frac{\langle F_S\rangle}{\langle S\rangle}=
\kappa_{\rm mess} \frac{g_{\rm mess}^2}{16\pi^2} \sqrt{m_{3/2}M_*},
 \end{eqnarray}
where $\kappa_{\rm mess}$ is supposed to be of $O(1)$ \cite{DN3}, and
$g_{\rm mess}$ is the gauge coupling constant for the U(1)$_{\rm
mess}$ gauge interaction.  In the messenger sector, there are also
${\bf 5}+{\bf \bar{5}}$ representation of SU(5), {\it i.e.}
SU(2)$_{\rm L}$ doublets ($L$ and $\bar{L}$) and SU(3)$_{\rm C}$
triplets ($Q$ and $\bar{Q}$).  These have the superpotential %
 \begin{eqnarray}
W_{\rm mess}=\lambda_2 SL\bar{L} + \lambda_3 SQ\bar{Q}.
 \end{eqnarray}
Once $S$ and $F_S$ acquire VEVs, the scalar components of $L$ and $\bar{L}$
($Q$ and $\bar{Q}$) have a mass matrix of the form
 \begin{eqnarray}
\left(\begin{array}{cc}
\lambda^2\langle S\rangle^2 & \lambda \langle F_S\rangle \\
\lambda \langle F_S\rangle  & \lambda^2\langle S\rangle^2\\
\end{array}\right),
 \end{eqnarray}
while the fermionic components have mass $\lambda\langle
S\rangle$.  Therefore, SUSY is broken in the mass spectrum of
vector-like ${\bf 5}+{\bf \bar{5}}$ messenger fields. 
By integrating out the messenger fields, the
soft SUSY breaking parameters are induced in the ordinary sector.
With $N_{\bf 5}$ pairs of vector-like ${\bf 5}+{\bf \bar{5}}$
multiplet, the gaugino masses, $m_{\rm G1}$, $m_{\rm G2}$, and $m_{\rm
G3}$ are given by
 \begin{eqnarray}
m_{{\rm G}i} = \frac{g_i^2}{16\pi^2}c_i\Lambda_{\rm mess} N_{\bf 5},
 \end{eqnarray}
where $c_3=c_2=1$, and $c_1=\frac{5}{3}$ in our convention.  On the
other hand, the masses for the sfermions $m_{\tilde{f}}$
($\tilde{f}=\tilde{u}_{\rm R},\tilde{d}_{\rm R},
\tilde{q}_{\rm L},\tilde{l}_{\rm L}$, and $\tilde{e}_{\rm R}$) are 
given by \cite{ACW}
 \begin{eqnarray}
m_{\tilde{f}}^2 = 2\Lambda_{\rm mess}^2N_{\bf 5}\sum_{i=1}^{3}
C_i \left(\frac{g_i^2}{16\pi^2}\right)^{2}.
 \end{eqnarray}
Here, $C_1=\frac{5}{3}Y^2$ with $Y$ being the usual hypercharge, and
$C_i=\frac{4}{3}$ and $\frac{3}{4}$ if $\tilde{f}$ is in the
fundamental representation of SU(3)$_{\rm C}$ and SU(2)$_{\rm L}$, and
$C_i=0$ for the gauge singlets.

Combining the above relations with the experimental bounds on the
sparticle masses, we can obtain a lower bound on $\Lambda_{\rm mess}$,
and hence the gravitino mass.  A lighter state gives us a more
stringent constraint.  For most parameters, the lightest among the
sfermions is the 
right handed selectron, whose mass $m_{\tilde{e}_{\rm R}}$ is
given by\footnote
{In fact, the selectron mass receives a correction from
renormalization effects. However, the correction is less than 10\%,
and the following arguments are almost unchanged even if we include
such effects.}
 \begin{eqnarray}
m_{\tilde{e}_{\rm R}}^2 \sim
\frac{10}{3} 
\left(\frac{g_1^2(\Lambda_{\rm mess})}{16\pi^2}\right)^2
\Lambda_{\rm mess}^2 N_{\bf 5}
-m_Z^2 \sin^2\theta_{\rm W}\cos 2\beta.
 \end{eqnarray}
For $\tan\beta$ close to 1, the right-handed selectron mass gives us 
a stringent bound on the messenger scale.  Even if we adopt a 
conservative constraint of $m_{\tilde{e}_{\rm R}}\geq 45{\rm 
GeV}$,\footnote{We are aware that LEP-II has placed stronger 
limits on the $m_{\tilde{e}}$, but as a function of the neutralino 
mass.  The mass difference between $m_{\tilde{e}}$ and the lightest 
neutralino $\sim \tilde{B}$ is not large in the LEGM models, and the 
constraint weakens substantially in this situation.  We chose the 
model-independent LEP bound because of this reason.} we obtain
 \begin{eqnarray}
\Lambda_{\rm mess}\gtrsim 2\times 10^4{\rm GeV}
\times \frac{1}{\sqrt{N_{\bf 5}}}.
\label{lambda_lower}
 \end{eqnarray}
If $\tan\beta$ is large, the $D$-term contribution enhances the right 
handed selectron mass, and $m_{\tilde{e}_{\rm R}}$ can be 
larger than the experimental limit with smaller value of $\Lambda_{\rm 
mess}$. In that case,  however, the sneutrino mass 
$m_{\tilde{\nu}_{\rm L}}$
 \begin{eqnarray}
m_{\tilde{\nu}_{\rm L}}^2 \sim
\left\{ \frac{3}{2}
\left(\frac{g_2^2(\Lambda_{\rm mess})}{16\pi^2}\right)^2
+ \frac{5}{6}
\left(\frac{g_1^2(\Lambda_{\rm mess})}{16\pi^2}\right)^2
\right\}
\Lambda_{\rm mess}^2N_{\bf 5}
+\frac{1}{2}m_Z^2 \cos 2\beta,
 \end{eqnarray}
receives a negative contribution from the $D$-term, and $\Lambda_{\rm 
mess}$ is still constrained to be larger than $\sim 2\times 10^4{\rm 
GeV}$.  Therefore, the bound (\ref{lambda_lower}) holds for all values of 
$\tan\beta$.  

In order to translate the above constraint (\ref{lambda_lower}) into
a lower bound on the gravitino mass, we need information about
the gauge coupling constant of the U(1)$_{\rm mess}$; larger $g_{\rm
mess}$ (at the messenger scale) gives us a less stringent constraint on
the gravitino mass.  However, if $g_{\rm mess}$ is too large at the
messenger scale, it blows up below the Planck scale or even below the
GUT scale.\footnote{There is the logical possibility of employing an
asymptotically free non-abelian gauge group as the messenger group, and
assume that its scale parameter is very close to $\Lambda_{\rm DSB}$.
We, however, consider it as an exponential fine-tuning of parameters.}
By using the 1-loop renormalization group equation, the
U(1)$_{\rm mess}$ gauge coupling constant at the messenger scale is
constrained by
 \begin{eqnarray}
\frac{g_{\rm mess}^2(\Lambda_{\rm mess})}{16\pi^2} \lesssim
\frac{1}{b_{\rm mess}
\ln (\Lambda_{\rm cut}^{2}/\Lambda_{\rm mess}^{2})},
 \end{eqnarray}
where $\Lambda_{\rm cut}$ is the cutoff scale where the perturbative 
picture may break down, and ${b_{\rm mess}}=\sum_A Q_{{\rm mess},A}^2$ 
is the sum of the squared charge of the messengers.  Typically, 
${b_{\rm mess}}\sim 10$.  (For the model proposed in \cite{DN2}, 
${b_{\rm mess}}=\frac{34}{3}$, and for a model given in \cite{DN3}, 
${b_{\rm mess}}=14$.)  Assuming ${b_{\rm mess}}=10$ and $\Lambda_{\rm 
cut}\sim M_{\rm GUT}$, $g_{\rm mess}$ at the messenger scale is 
constrained to be smaller than $\sim 0.5$, and
 \begin{eqnarray}
m_{3/2}\gtrsim
70 {\rm keV}\times
\frac{\kappa_{\rm mess}^{-2}}{N_{\bf 5}}
\left(\frac{g_{\rm mess}}{0.5}\right)^{-2}
\left(\frac{m_{\tilde{e}_{\rm R}}}{\rm 45GeV}\right)^2.
\label{mgravitino_min}
 \end{eqnarray}
In the minimal model, $N_{\bf 5}=1$, and if we assume the perturbative
unification of the gauge coupling constants in the MSSM, $N_{\bf 5}\leq 4$
\cite{CM}. Therefore, in any case, the lower bound above is about one or
two order of magnitude larger than the cosmological upper bound
(\ref{1kev}).  Notice that the lower bound on the gravitino mass
increases as the experimental lower bound on the sparticle masses
increases.\footnote{A collorary to this analysis is that it is unlikely
to have a decay of a sparticle into the gravitino inside a collider
detector.  This casts some doubts on the naturalness of $ll\gamma\gamma$
signature at CDF in the LEGM models.  A possible way out is to employ the
vector-like model \cite{IY} and couple a singlet field directly to the
messenger fields in the superpotential \cite{IT}.  This model, however,
probably suffers from a tunneling to a color- and charge-breaking
supersymmetric minimum if all coupling constants are $O(1)$ \cite{DDR}.}

Based on the above estimations, we define the canonical set of the
parameters for our following analysis:
 \begin{eqnarray}
&&m_{3/2} = 100{\rm keV},\\
&&\Lambda_{\rm mess} = 3\times 10^{4} {\rm GeV},\\
&&\langle F_S\rangle^{1/2}= \langle S\rangle = 
\Lambda_{\rm mess}.
 \end{eqnarray}
Note that it is easy to raise the gravitino mass; we only have to
assume a smaller value for the gauge coupling constant for the
U(1)$_{\rm mess}$.  In the following analysis, we basically
assume the above set of parameters, and we also discuss how our
results change if we vary them.

The above estimations are based on perturbative calculations, and
one may worry that a strong coupling in the dynamical sector may allow
us to lower the gravitino mass. Such a scenario seems unlikely,
however. To see this, it is convenient to define the ``vacuum
polarizations'' from the DSB sector for the U(1)$_{\rm mess}$ gauge
multiplet:
 \begin{eqnarray}
&&{\rm F.T.}\left\langle 0| T (A_\mu A_\nu) |0\right\rangle_{\rm 1PI}
= i q^2 \Pi_A (q^2)g_{\mu\nu} ,
\\
&&{\rm F.T.}\left\langle 0| T (\lambda \lambda ) 
|0\right\rangle_{\rm 1PI}
= (-i)\{ \not\!{q}\Pi_\lambda (q^2) + \Sigma(q^2) \},
\\
&&{\rm F.T.}\left\langle 0|T (D D) |0\right\rangle_{\rm 1PI}
= i\Pi_D(q^2),
 \end{eqnarray}
where F.T. stands for the four-dimensional Fourier Transform to the 
momentum space, and 1PI for one-particle irreducible diagrams.
At tree level, $\Pi_A=\Pi_\lambda =\Pi_D = \Sigma =0$.  
These quantities are radiative corrections of $O(g_{\rm 
mess}^2/16\pi^2)$ if the perturbative calculation is reliable.  The 
messenger scale in the LEGM model is induced by integrating out the 
SUSY breaking sector and the U(1)$_{\rm mess}$ gauge 
multiplet. 
By using $\Pi_A$, $\Pi_\lambda$ and $\Pi_D$, the SUSY breaking 
scalar mass in the messenger sector ($\sim$ the messenger scale) is 
given by
 \begin{eqnarray}
m_{\rm mess}^2\sim g_{\rm mess}^2
\int\frac{d^4q}{(2\pi)^4i}\frac{1}{q^2}
\{ 3\Pi_A (q^2) -4\Pi_\lambda (q^2) +\Pi_D (q^2) \}.
\label{int_for_messenger}
 \end{eqnarray}
If we limit ourselves to the physics of the DSB sector, there is no
pole in the $\Pi$ functions at $q^{2} = 0$ which, if present, implies the
Higgs mechanism for the U(1)$_{\rm mess}$ gauge group.  The only
singularities in $\Pi$ functions are, therefore, branch cuts which
appear above certain threshold $q^{2} \gtrsim \Lambda_{\rm DSB}^{2}$
which is the only scale in the problem.  Then the integrations in
Eq.~(\ref{int_for_messenger}) can be Wick rotated and we obtain
 \begin{eqnarray}
m_{\rm mess}^2\sim - \frac{g_{\rm mess}^2}{16\pi^2}
\int_{0}^{\infty} d q_{E}^{2} 
\{ 3\Pi_A (-q_{E}^2) -4\Pi_\lambda (-q_{E}^2) +\Pi_D (-q_{E}^2) \}.
\label{int_for_messenger2}
 \end{eqnarray}
Now it is clear that the integration is purely Euclidean, and hence 
all $\Pi$ functions are far off-shell.  Thus, the perturbative result 
is essentially reliable even when the DSB sector is strongly 
coupled.  It is also useful to recall that similar calculations of 
vacuum polarization amplitudes in QCD tend to agree with lowest order 
perturbative results for the running of the fine-structure constant, or 
the scaled-up QCD estimate of the electroweak $S$-parameter.\footnote
{Note that the $S$-parameter is defined by the 
vacuum polarization amplitudes at $q^{2} = 0$, and hence more sensitive 
to the non-perturbative effects than our case 
(\ref{int_for_messenger2}) which smears them over the wide range of 
$q^{2}$.  Still, a perturbative estimate of $S$ differs from the scaled-up 
QCD only by a factor of two.  One may also estimate the $S$-parameter 
by assuming that it is dominated by the $\rho$ and $a_{1}$ poles.  Then 
the result is obtained by the tree-level process.  Even so, the 
coupling of the resonances to the current operator has a factor of 
$1/4\pi$ and the counting of $1/4\pi$ factors remains the same in as the 
perturbative one-loop result.  HM thanks Christopher D. Carone on this 
point.}
We therefore conclude that there is no significant enhancement of the 
resulting $m_{\rm mess}^{2}$, and hence the estimates of the messenger
scale and the resulting gravitino mass (\ref{mgravitino_min}) can be
trusted.

The constraint (\ref{mgravitino_min}) sets severe bounds on
cosmology. In particular, we need some mechanism to generate a
dilution factor of $\sim (m_{3/2}/2h^2{\rm keV})$ at a relatively low
temperature below the upper bound on the maximum temperature given in 
Eq.~(\ref{Tmax}) and in Fig.~\ref{fig:tr_max}, if the gravitino mass is larger than
$2h^2{\rm keV}$. 
Furthermore, even if we adopt such a large entropy production at a low
temperature, baryogenesis may still be a
problem. The Affleck-Dine mechanism \cite{AD} 
for baryogenesis is one of the
possibilities to generate baryons at a relatively low
temperature.\footnote{
Electroweak baryogenesis may be another possibility to generate
baryon asymmetry at relatively low temperatures. However, the resulting
baryon-to-photon ratio depends on the details of the complicated 
dynamics of the phase transition. Furthermore, the generated baryon
asymmetry would not be large, if any, and would probably not survive a
huge entropy production to dilute the string moduli fields as discussed
in Section~\ref{sec:moduli}.}

However, in the LEGM models, the behavior of the flat direction at large
amplitude is quite different from the usual supergravity case.  Thus,
even if we assume the Affleck-Dine mechanism, it is a
non-trivial question whether we can have enough baryon number density.
In the following sections, we pursue this possibility, and as a
result, we will see that the Affleck-Dine mechanism works sufficiently
well, enough to explain the present value of the baryon-to-entropy
ratio.

\section{Flat Directions in the LEGM Models}
\label{sec:flat-dir}
\setcounter{equation}{0}
\setcounter{footnote}{0}

As discussed in the previous sections, the constraint from the 
gravitino cosmology is quite severe in models with the LEGM. 
Therefore, it is preferable to look for baryogenesis scenarios which 
do not require high temperatures.

We will focus on the Affleck--Dine baryogenesis in
the LEGM models in this paper.  
There is one crucial difference from the hidden sector models: the 
potential along the MSSM flat directions is not simply parabolic. 
Therefore we discuss the form of the potential 
first in this section, and then its implication to the Affleck--Dine
baryogenesis in the next section.

In the hidden sector models where the supersymmetry breaking effect is 
mediated by Planck-scale operators, the soft supersymmetry 
breaking parameters are actually ``hard'', in the sense that they 
renormalize as usual mass terms between the Planck-scale and the weak 
scale.  On the other hand, the supersymmetry breaking scalar mass 
terms are suppressed beyond the messenger scale in the LEGM models.

This is analogous to the situation in the QCD. The current masses of
the quarks renormalize according to the ordinary perturbation theory.
They are ``hard'' masses.  However the constituent quark masses are
suppressed by a power of the energy: ``soft''.  This is because the
constituent quark masses are dynamically generated by the spontaneous
chiral symmetry breaking, which is characterized by the order
parameter $\langle \bar{q} q \rangle$.  The constituent quark mass has
to be proportional to this order parameter.  At high momentum
transfer, a dimensional analysis then tells us that the effective
constituent mass behaves as $m_{\rm const} (Q^{2}) \sim \langle
\bar{q} q \rangle/Q^{2}$.

The same argument applies to the soft supersymmetry breaking masses 
from the LEGM.  Supersymmetry is broken by 
an $F$-component of a chiral superfield, $\langle F_{S} \rangle \neq 
0$.  The soft supersymmetry breaking scalar mass is necessarily 
proportional to the order parameter of supersymmetry breaking, {\it 
i.e.}\/, $m^{2} \propto \langle F_{S}\rangle^{\dagger} \langle F_{S} 
\rangle$.  A dimensional analysis tells us that it is suppressed at 
high momentum transfers, $m^{2} (Q^{2}) \sim \langle 
F_{S}\rangle^{\dagger} \langle F_{S} \rangle / Q^{2}$.  Therefore, 
the supersymmetry breaking parameters ``shut off'' at high energies.

The potential of a MSSM flat direction is given simply by $V = m^{2}
|\phi|^{2}$, where $m^{2}$ is a soft supersymmetry breaking mass.  A
renormalization group improvement gives us $V = m^{2} (|\phi|^{2})
|\phi|^{2}$.  In the hidden sector case, $m^{2} (|\phi|^{2})$ has only
a logarithmic dependence on $|\phi|^{2}$ and hence can be taken
approximately constant unless it crosses zero at some energy scale.
For most cosmological applications, this is a sufficiently good
description.  In the LEGM models, however, the effective mass $m^{2}
(|\phi|^{2})$ exhibits a power dependence on $|\phi|^{2}$ which cannot
be neglected.  We expect that $m^{2} (|\phi|^{2})$ behaves as $\langle
F_{S} \rangle^{\dagger} \langle F_{S} \rangle/|\phi|^{2}$ for large
$|\phi|$, and hence the potential behaves approximately like a constant
for $|\phi| > \langle S \rangle$, which is the mass scale of the
messengers.

We performed an explicit two-loop calculation of the effective
potential $V(\phi)$ and its details are presented in
Appendix~\ref{app:potential}.  Here we only quote the result.  As
expected, the potential behaves parabolically around the origin, while
it becomes approximately constant for large $|\phi|$; actually it keeps
growing slowly as $(\ln |\phi|^{2})^{2}$.  The potential of a MSSM flat
direction behaves as
\begin{equation}
	V (\phi)  \sim
		\left( \frac{g^2}{(4\pi)^2} \right)^{2}
		\left( \frac{\langle F_{S} \rangle}{\langle S \rangle}
			\right)^{2} |\phi|^{2} + O(|\phi|^{4})
\end{equation}
for small $|\phi| \ll \langle S \rangle$, and
\begin{equation}
	V (\phi)  \sim V_0
		\left(\ln \frac{|\phi|^{2}}{\langle S \rangle^{2}}
			\right)^{2}
	\label{Vphi-large}
\end{equation}
 with
 \begin{eqnarray}
	V_0\sim \frac{g^2}{(4\pi)^4} \langle F_{S} \rangle^{2}
   	\sim (3 \times 10^{3}{\rm GeV})^4,
   	\label{V0}
 \end{eqnarray}
for large $|\phi| \gg \langle S \rangle$.  Here, $g$ generically refers
to standard model gauge coupling constants.\footnote{
The readers may wonder why Eqs.~(\ref{Vphi-large}) and (\ref{V0}) 
have only two powers of gauge coupling constants despite the 
two-loop-ness of the effective potential.  This is the result of an 
explicit calculation, and we can also explain it in a simple way.  
When the field value is large, the standard model gauge multiplets 
acquire large masses of order $g\phi$.  The effective potential is 
generated by the exchange of heavy gauge multiplets, and hence it is
suppressed by $1/|g\phi|^{2}$.  This cancels two powers in gauge 
coupling constants.}

For extremely large $|\phi|$, however, the contribution from 
supergravity becomes important.  Supergravity generates a 
contribution to the scalar potentials $\sim m_{3/2}^{2} |\phi|^{2}$ 
for any $|\phi|$.\footnote
{This is true for the minimal supergravity and its variants.  
This contribution, however, is not there in no-scale supergravity, or 
in general, supergravity with Heisenberg symmetry \cite{MSYY,GMO}.  In 
such a case, $\phi_{\rm eq}$ must be taken at $M_{*}$ in the rest of 
the discussions.  } 
To determine the relative importance of the LEGM and supergravity 
contributions, we compare their derivative $V'$ because this is the 
quantity which appears in the equation of motion.  The derivative of 
the potential from the LEGM is
\begin{equation}
	\frac{\partial V}{\partial \phi} 
	\sim V_{0}\frac{2 \phi^{*}}{|\phi|^{2}} 
		\left( \ln \frac{|\phi|^{2}}{\langle S \rangle^{2}}
			\right) ,
\end{equation}
which is to be compared with the supergravity contribution $\partial 
V/\partial \phi = m_{3/2}^{2} \phi^{*}$.  The supergravity 
contribution is more important above a threshold value $\phi_{\rm eq}$ 
which is given by
\begin{equation}
	\phi_{\rm eq} \sim 
		\left\{ \frac{2 V_{0}}{m_{3/2}^{2}} 
		\left( \ln \frac{|\phi_{\rm eq}|^{2}}
		{\langle S \rangle^{2}}
		\right) \right\}^{1/2}
		\sim 7\times 10^{11}~\mbox{GeV} \times 
		\left(\frac{m_{3/2}}{\rm 100 keV}\right)^{-1}
		\left(\frac{V_0^{1/4}}{3 \times 10^3~\mbox{GeV}}
 		\right)^2.
\end{equation}
The motion of the flat direction is determined by the effective 
potential given in this section and the canonical kinetic term, and 
there is no need to include the wave function renormalization factor 
at this order in perturbation theory.  See Appendix \ref{app:wavefn} 
for details.

\section{Affleck--Dine Baryogenesis in the LEGM Models}
\label{sec:ADinLEGM}
\setcounter{equation}{0}
\setcounter{footnote}{0}

In this section, we aim at estimating the size of baryon-to-entropy
ratio from Affleck--Dine baryogenesis in the LEGM models.  Because of the
multiple scales in the problem, the discussion becomes somewhat
complicated.  The basic conclusion is that the Affleck--Dine
baryogenesis works as efficiently as in the hidden sector case, but in a
much more non-trivial manner.  Finally we discuss the possible dilution
of gravitinos via the decay of the Affleck--Dine flat direction, and find
that the gravitinos can be diluted below the closure limit if the
initial amplitude of the flat direction is sufficiently large.

\subsection{Generalities}

In Affleck--Dine baryogenesis \cite{AD}, one assumes that a MSSM flat
direction has a large amplitude at the end of the primordial
inflation.  The mechanism to achieve a large amplitude varies: a
negative curvature from a non-minimal K\"ahler potential \cite{DRT}, or
no-scale supergravity \cite{GMO}.  In any case, it tends to be equal
to or larger than the expansion rate of the Universe during inflation
$H_{\rm inf}\sim 10^{11}$~GeV -- $10^{13}$~GeV depending on
inflationary scenarios.  We phenomenologically parameterize it just as
the initial amplitude $\phi_{0}$.

A typical assumption is that there is a baryon-number violating K\"ahler
potential term $K \sim l^{*} q^{*} u^{c} d^{c}/M_{*}^{2}$, where $M_{*}$
is the reduced Planck scale.\footnote{It could also well be the
GUT-scale $M_{\rm GUT}$. However, in this paper, we assume this form of
the baryon number violating operator for simplicity. The extension to
the case with $K\sim l^{*}q^{*}u^{c}d^{c}/M_{\rm GUT}^{2}$ is trivial,
and one can easily estimate the resulting baryon-to-entropy ratio.}  The
supersymmetry breaking effects from the LEGM generates a term in the
potential\footnote{We have not calculated this explicitly.  This form is
expected based on the analogy to the calculation of the effective
potential in the previous section.  The only difference is that the 
previous one arises from the kinetic term $\phi^* \phi$ in the K\"ahler
potential rather than from a non-renormalizable term 
$l^{*}q^{*}u^{c}d^{c}/M_{*}^{2}$ here.}
\begin{eqnarray}
{\cal O} &\sim& \int d^4 \theta \frac{g^4}{(4\pi)^4} 
	\frac{|\theta^2 F_S|^2}{g^{2}|\phi|^2}
	\left(\ln \frac{|\phi|^{2}}{\langle S \rangle^{2}} \right)^{2}
	\frac{1}{M_*^2} l^{*}q^{*}u^{c}d^{c} + h.c.
\nonumber \\
	&=& \frac{V_0}{|\phi|^{2}} 
	\left(\ln \frac{|\phi|^{2}}{\langle S \rangle^{2}} \right)^{2}
	\frac{1}{M_*^2} 
	\tilde{l}^{*} \tilde{q}^{*} \tilde{u}^{c} \tilde{d}^{c} + h.c.,
\end{eqnarray}
while the supergravity effect induces an operator
in the scalar potential
 \begin{equation}
	{\cal O} \sim 
	\int d^4 \theta \frac{|\theta^2 F_0|^2}{M_*^2}
	\frac{1}{M_*^2} l^{*}q^{*}u^{c}d^{c} + h.c.
	=
	\frac{m_{3/2}^{2}}{M_{*}^{2}} 
	\tilde{l}^{*} \tilde{q}^{*} \tilde{u}^{c} \tilde{d}^{c} 
	+ h.c.
\end{equation}
The LEGM operator is dominant if $|\phi_0| \lesssim \phi_{\rm eq}$, while
the supergravity one dominates if $|\phi_0| \gtrsim \phi_{\rm eq}$.
Therefore, we discuss the two cases separately below.  In either case, the
size of the baryon-number violating operator is much smaller than in the
hidden sector case (Appendix \ref{app:hidden}).  
It turns out, however, that the
baryogenesis proceeds efficiently with these operators.

Below we generically refer to the fields as $\phi$ without
distinguishing various species. The baryon number in the 
scalar sector is given by
 \begin{eqnarray}
n_B = i(\dot{\phi}^* \phi - \dot{\phi} \phi^*),
\label{n_B}
 \end{eqnarray}
while the baryon number violating operator is written as
 \begin{equation}
{\cal O} \sim \left[m_{3/2}^{2} +
		\frac{V_{0}}{|\phi|^{2}} 
		\left(\ln \frac{|\phi|^{2}}{\langle S \rangle^{2}} 
			\right)^{2} \right]
	\frac{1}{M_{*}^{2}}(\phi^4 + \phi^{*4}).
\label{O_Bviolate}
 \end{equation}

\subsection{$|\phi_0| \gtrsim \phi_{\rm eq}$}

For sufficiently large $|\phi_{0}|$ ($|\phi_{0}|\gtrsim \phi_{\rm eq}$),
the supergravity contribution is initially important, and the field
begins to roll down the potential when the expansion rate of the
Universe $H$ is comparable to $H \sim m_{3/2}$.  Let us first estimate
the primordial baryon number asymmetry in this case, $|\phi_0|\gtrsim
\phi_{\rm eq}$.

A rough estimation of the baryon asymmetry, which is generated just
after the start of the oscillation of the $\phi$ field can be done
only by using simple order of magnitude arguments.  With the above baryon number
violating operator, the time evolution of the baryon number is given
by
 \begin{equation}
	\dot{n}_{B} + 3 H n_{B} = i\left(
		\frac{\partial {\cal O}}{\partial \phi} \phi
		- \frac{\partial {\cal O}}{\partial \phi^{*}}
                \phi^{*}\right) .
	\label{nBeom}
 \end{equation}
When the field begins to roll down the potential, its initial motion 
is slow, and one can neglect the $\dot{n}_{B}$ term in the equation (see 
Appendix~\ref{app:linde}).  Then the resulting baryon number can be 
estimated by \cite{Linde}
 \begin{equation}
	n_{B} \sim \frac{i}{3H} \left(
		\frac{\partial {\cal O}}{\partial \phi} \phi
		- \frac{\partial {\cal O}}{\partial \phi^{*}} 
		\phi^{*}
		\right) .
\label{nb=nbdot/H}
 \end{equation}
Using the approximate order of magnitude of the operator and $H\sim
m_{3/2}$, we obtain
 \begin{equation}
	n_{B} \sim 
		\frac{m_{3/2} {\rm Im}(\phi_{0}^{4})}{M_{*}^{2}}.
 \end{equation}
It depends on the imaginary part of the initial amplitude.  The 
entropy of the radiation at this stage is given by $s \sim g_* T^{3}$ 
while the energy density $\rho_{\rm rad} 
\sim g_* T^{4} \sim m_{3/2}^{2} M_{*}^{2}$.
By putting them together, we can estimate the baryon-to-entropy ratio,
 \begin{equation}
	\frac{n_{B}}{s} \sim g_*^{-1/4}
		\frac{{\rm Im}(\phi_{0}^{4})}{m_{3/2}^{1/2} M_{*}^{7/2}}
	\sim 4\times 10^{10} \times \left(\frac{|\phi_0|}{M_*}\right)^4
	\left(\frac{m_{3/2}}{\rm 100keV}\right)^{-1/2}\sin 4\theta_0,
 \label{nb/s} 
 \end{equation} 
where the initial amplitude is parameterized as
$\phi_0=|\phi_0|e^{i\theta_0}$. As one can see, a large baryon asymmetry
can be generated, if the initial amplitude of $\phi$ is not too small.
Therefore, a large enough baryon number can remain in
this scenario, even if there is a substantial entropy production after the
Affleck-Dine baryogenesis. Note that one obtains exactly the same
expression in the 
hidden sector models, but with a different $m_{3/2}$.

The present baryon-to-entropy ratio is also given by the above
formula, if there is no significant entropy production. However, in
a realistic situation, there can be entropy production. In
particular, the decay of the Affleck-Dine field $\phi$ may produce a
large amount of entropy. Furthermore, for $m_{3/2}\gtrsim 2h^2$ keV,
we need a non-negligible entropy production to dilute the primordial
gravitino. If there is an entropy production after the Affleck-Dine
baryogenesis, the primordial baryon number density is also diluted. In
the following, we discuss how an entropy production affects the
results.

We estimate the entropy production due to the decay of the flat
direction. As discussed in the previous section, $\phi$ starts to
oscillate when $T=T_0\sim g_*^{-1/4}\sqrt{m_{3/2}M_*}$, if
$|\phi_0|\gtrsim \phi_{\rm eq}$. During $|\phi|\gtrsim\phi_{\rm eq}$,
the potential for $\phi$ is dominated by the supergravity contribution,
and hence
 \begin{eqnarray}
|\phi|^2 R^3 = {\rm const.} ~~~ 
(\phi_{\rm eq}\lesssim|\phi|\lesssim |\phi_0|).
\label{phi^2R^3=const}
 \end{eqnarray}
Thus, the temperature at $|\phi|\sim\phi_{\rm eq}$, which we denote
$T_{\rm eq}$, is estimated as
 \begin{eqnarray}
T_{\rm eq} \sim T_0 \left(\frac{\phi_{\rm eq}}{|\phi_0|}\right)^{2/3}.
 \end{eqnarray}
For $|\phi|\lesssim\phi_{\rm eq}$, the potential for the flat direction
is dominated by the LEGM piece, and the evolution for $\phi$ does not
obey the relation (\ref{phi^2R^3=const}). By using the virial theorem,
the evolution of the flat direction can be estimated, and is given by
 \begin{eqnarray}
|\phi| R^3 = {\rm const.} ~~~ 
(\langle S \rangle \lesssim |\phi| \lesssim\phi_{\rm eq}).
\label{phiR^3=const}
 \end{eqnarray}
(See Appendix \ref{app:virial}.) By using the above relations, we
obtain the dilution factor due to the decay of the flat direction.

Now an important question is at what field amplitude $\phi$ decays 
into radiation.  When the motion is dominated by a parabolic term, we 
know the time dependence of the oscillation (just a harmonic 
oscillator), and we can calculate the rate of particle production in 
such a background.  The result is known to be the same as the single 
particle decay rate, if the amplitude is not too large compared to the 
oscillation frequency.  Once the amplitude becomes comparable to 
$\langle S\rangle$, the potential is almost parabolic, and we find 
that the coherent oscillation decays into radiation rapidly.  On the 
other hand, a corresponding calculation is difficult when the 
potential is dominated by the logarithmic term.  In our analysis, we 
regard the decay amplitude, $\phi_{\rm dec}$, as a free parameter, and 
discuss the $\phi_{\rm dec}$-dependence of the results.\footnote{For a 
canonical parameter, we take $\phi_{\rm dec} \sim 10^5$~GeV for 
estimating dilution factors.  Since $\phi$ decays at $\phi \sim 
\langle S \rangle \sim 3\times 10^{4}$~GeV at latest, this choice 
gives us the minimum estimate of the baryon asymmetry.  If $\phi$ 
decays earlier, the dilution factor is less and the baryon asymmetry 
is larger.  We likely overestimate the dilution factor with this 
choice.  We will come back to this point later when we discuss a 
possible dilution of gravitinos from the decay of the flat direction.  }

The temperature of the background radiation at the decay time of the
flat direction, $T_{\rm dec}$, is given by
 \begin{eqnarray} 
T_{\rm dec}\sim
T_{\rm eq}\left(\frac{\phi_{\rm dec}}{\phi_{\rm eq}}\right)^{1/3} 
\sim
T_0\left(\frac{\phi_{\rm dec}}{\phi_{\rm eq}}\right)^{1/3}
\left(\frac{\phi_{\rm eq}}{|\phi_0|}\right)^{2/3}.
 \end{eqnarray}
On the contrary, the energy density of the flat direction is
 \begin{eqnarray} 
\rho_{\rm flat} \sim V_0.
 \end{eqnarray}
Then, if $\rho_{\rm flat}\gtrsim\rho_{\rm rad}$, the 
dilution factor from the decay is given by
 \begin{eqnarray}
D
& \sim&
\left(\frac{\rho_{\rm flat}}{\rho_{\rm rad}} \right)^{3/4}
\sim \left( \frac{V_0}{g_*T_{\rm dec}^4} \right)^{3/4}
\nonumber \\
& \sim & 6\times 10^8
\left(\frac{|\phi_0|}{M_*}\right)^2
\left(\frac{\phi_{\rm dec}}{\rm 10^5GeV}\right)^{-1}
\left(\frac{m_{3/2}}{\rm 100keV}\right)^{-1/2}
\left(\frac{V_0^{1/4}}{3 \times 10^3~\mbox{GeV}}\right).
\label{dilution-factor}
 \end{eqnarray}
Note that $D\sim 1$ if $\rho_{\rm flat}\lesssim\rho_{\rm rad}$,
or in terms of the initial amplitude, $\phi_{0} \lesssim 
10^{14}$~GeV.

Combining the above dilution factor with the estimation of the
primordial baryon number density given in Eq.~(\ref{nb/s}), we
obtain the present baryon number asymmetry. In order to make a
pessimistic estimate of the resulting baryon asymmetry, we assume that
the flat direction decays only when its amplitude is as small as
$\langle S\rangle$.  This assumption maximizes the entropy production,
and hence, gives us the minimum value for the baryon asymmetry.  If it
decays earlier, then the entropy production is less and hence the
baryon asymmetry is larger.  With this caveat in mind, we can make an
estimate of the resulting baryon-to-entropy ratio, and in the case
with entropy production ($D>1$), the resulting baryon-to-entropy ratio
is given by
 \begin{equation}
	\frac{n_{B}}{s} \sim 
	D^{-1} g_*^{-1/4}
	\frac{{\rm Im}(\phi_{0}^{4})}{m_{3/2}^{1/2} M_{*}^{7/2}}
	\sim 70
	\times \left(\frac{|\phi_0|}{M_*}\right)^2
	\left(\frac{\phi_{\rm dec}}{\rm 10^5~GeV}\right)
 	\left(\frac{V_0^{1/4}}{3 \times 10^3~\mbox{GeV}}\right)
	\sin 4\theta_0.
 \label{nb/s/D}
 \end{equation}
Note that the result is independent of the gravitino mass.
It is intriguing that the final result is more or less the same as 
in the hidden sector case Eq.~(\ref{nB/s-hidden}).

\subsection{$|\phi_0| \lesssim \phi_{\rm eq}$}

Next, we discuss the case of $|\phi_0|\lesssim \phi_{\rm
eq}$.\footnote{As noted before, the no-scale supergravity does not
generate potential term proportional to $m_{3/2}^{2}$ and hence the
evolution of the flat direction is always dominated by the LEGM
piece.  Then the formulae presented in this subsection must be used
even for a larger $|\phi_{0}|\sim M_{*}$. Such a large $|\phi_{0}|$ is
indeed expected in the no-scale case because the flat directions
remain flat even during the inflation \cite{GMO}.} In this
case, the potential for the flat direction is dominated by the
LEGM-piece, and the flat direction starts to move when
$H\sim\sqrt{|V' (\phi_{0})|/|\phi_{0}|}\sim 
\sqrt{V_0 (\ln |\phi_{0}|^2/\langle S
\rangle^2)}/|\phi_0|$.  The temperature at this
stage, $T_0$, is estimated as
 \begin{eqnarray}
 T_0^2\sim g_*^{-1/2} M_* \frac{V_0^{1/2}}{|\phi_0|} 
 \left(\ln
 \frac{|\phi_{0}|^2}{\langle S \rangle^2} \right)^{1/2}.
 \end{eqnarray}
 Then, by using Eq.(\ref{nb=nbdot/H}), we can estimate the resulting
baryon number density, and hence the baryon-to-entropy ratio. 
Note that the baryon-number-violating operator ${\cal O}$ is different
from the previous case.  

Following exactly the same steps as in the previous case, we find
\begin{equation}
	n_{B} \sim 
		\frac{V_0^{1/2} {\rm Im}(\phi_{0}^{4})}
		{M_*^2 |\phi_{0}|}  \left(\ln
		 \frac{|\phi_{0}|^2}{\langle S \rangle^2} \right)^{3/2} 
	= \frac{V_0^{1/2} |\phi_{0}|^3}{M_*^2}
	 	\left(\ln
 		\frac{|\phi_{0}|^2}{\langle S \rangle^2} \right)^{3/2} 
 		\sin 4 \theta_0,
\end{equation}
when the flat direction starts to move.
Therefore, the baryon-to-entropy ratio is given by
\begin{eqnarray}
	\frac{n_{B}}{s} & \sim&  g_*^{-1/4}
		\frac{|\phi_{0}|^{9/2}}
		{V_0^{1/4} M_*^{7/2}}
		\left(\ln
 		\frac{|\phi_{0}|^2}{\langle S \rangle^2} \right)^{3/4}
		\sin 4\theta_0 .
		\nonumber \\
		& \sim & 6 \times 10^{-14} \times
			\left( \frac{|\phi_{0}|}{10^{12}~\mbox{GeV}} \right)^{9/2}
			\left( \frac{V_{0}^{1/4}}{3\times 10^{3}~\mbox{GeV}}
				\right)^{-1}
			\sin 4\theta_{0}
\label{nb/s-small}
\end{eqnarray}
which is typically too small.  It is useful to note that the above 
formula is larger by a factor of the logarithm than the corresponding 
formula Eq.~(\ref{nb/s}) for the case $\phi_{0} \gtrsim \phi_{\rm eq}$ 
when we substitute $\phi_{0} = \phi_{\rm eq}$.  Of course such a 
discontinuity cannot exist.  It simply means that there is a 
transition region at $\phi_{0} \sim \phi_{\rm eq}$ where there is a slight 
rise in $n_{B}/s$ when we cross $\phi_{0} \sim \phi_{\rm eq}$ from 
above.

The dilution factor can be estimated also along the lines of the previous
case.  We have
\begin{equation}
	T_{\rm dec} \sim 
	T_0 \left( \frac{\phi_{\rm dec}}{\phi_0} \right)^{1/3},
\end{equation}
and hence (if $\rho_{\rm flat} > \rho_{\rm rad}$),
 \begin{eqnarray}
D
&\sim&
\left(\frac{\rho_{\rm flat}}{\rho_{\rm rad}} \right)^{3/4}
\sim \left( \frac{V_0}{g_*T_{\rm dec}^4} \right)^{3/4}
\sim \frac{|\phi_0|^{5/2}}{M_*^{3/2} \phi_{\rm dec} (\ln
|\phi_0|^2/\langle S \rangle^2)^{3/4}}
\nonumber \\
&\sim& 2 \times 10^{-4} 
	\left( \frac{|\phi_0|}{10^{12}~\mbox{GeV}} \right)^{5/2}
	\left(\frac{\phi_{\rm dec}}{10^5~\mbox{GeV}}\right)^{-1}.
\end{eqnarray}
(More correctly, the dilution factor is $D = ((\rho_{\rm flat} + 
\rho_{\rm rad})/\rho_{\rm rad})^{3/4}$ and cannot be less than unity.)
Therefore, the dilution factor is much less important than in the
previous case.

\subsection{Numerical Analysis}
\begin{figure}
\centerline{\epsfxsize=0.8\textwidth \epsfbox{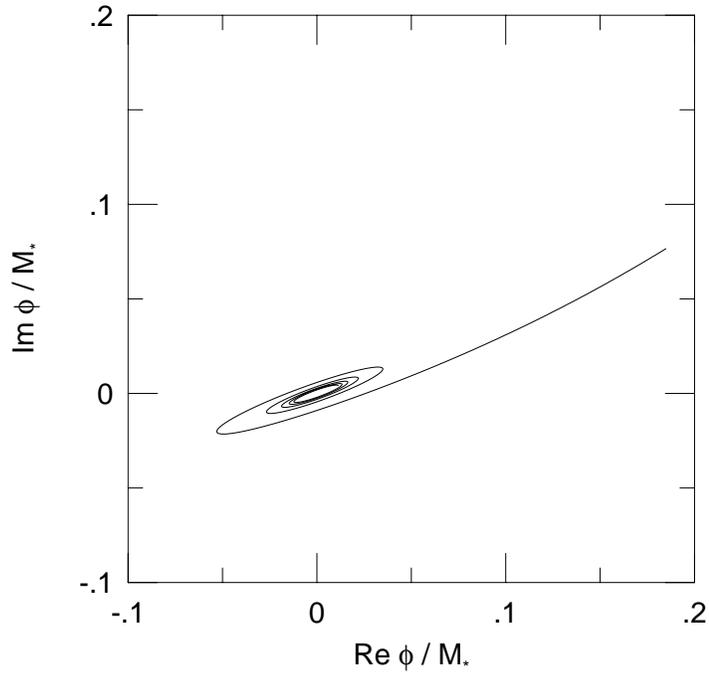}}
\caption{The initial motion of the flat direction with the potential
given in Eq. (\protect\ref{V_tot}). Here, we take $m_{3/2}=100$ keV,
and $\phi_0=0.2M_*e^{i\pi/8}$.}
\label{fig:motion}
\end{figure}

A more detailed behavior of the baryon-to-entropy ratio can be studied
by a numerical calculation. Here, to see the behavior of the
results, we show the resulting baryon-to-entropy ratio for a particular
set of parameters: $V_0=(3 \times 10^{3}{\rm GeV})^4$, 
$\langle S\rangle =3\times 10^4$~GeV, 
$\phi_{\rm dec}=10^5$ GeV, and $\sin 4\theta_0=1$. The results
for other sets of parameters can be easily estimated by using
Eqs.(\ref{nb/s}) and (\ref{nb/s/D}).  First, we solved the equation of
motion for the flat direction,
 \begin{eqnarray}
\ddot{\phi} + 3H\dot{\phi} + \frac{\partial V}{\partial \phi^*}
=0,
 \end{eqnarray}
with the potential\footnote
{The potential given in Eq.(\ref{V_tot}) is unbounded-below for
$\phi\gtrsim M_*$, and higher-dimension
operators are supposed to stabilize it. However, we
only consider the initial amplitude to be less than $\sim M_*$ in our
analysis, and hence the postulated potential is a good enough
approximation.} 
 \begin{eqnarray}
V = m_{3/2}^2|\phi|^2 
+ V_0\left(\ln\frac{|\phi|^2}{\langle S\rangle^2}\right)^2
-\left[m_{3/2}^{2} +
		\frac{V_{0}}{|\phi|^{2}} 
		\left(\ln \frac{|\phi|^{2}}{\langle S \rangle^{2}} 
			\right)^{2} \right]
	\frac{1}{M_{*}^{2}}(\phi^4 + \phi^{*4}).
\label{V_tot}
 \end{eqnarray} 
 We start our calculation at a temperature much higher than $T_0$
($T=10T_0$), and follow the evolution of the flat direction as well as
the temperature of the thermal bath. With the initial value
$\phi_0=0.2M_*e^{i\pi/8}$, the initial motion is shown in
Fig.~\ref{fig:motion}. As one can see, $\phi$ starts an elliptical
motion due to the baryon-number violating term in the potential. This
means that a non-vanishing baryon number is generated once $\phi$ starts to
oscillate. We indeed found the generated baryon number to be 
consistent with our estimates in the previous subsections within a 
factor of a few.

With the motion of the flat direction, we can calculate the
baryon number by using Eq.(\ref{n_B}), and hence the baryon-to-entropy
ratio.  After several cycles of oscillation, the baryon-to-entropy
ratio becomes almost constant. Then, the evolution of $\phi$ and $T$
can be easily traced by using Eqs.(\ref{phi^2R^3=const}) and
(\ref{phiR^3=const}) with $RT=$ const. Finally, we calculate the
dilution factor, $D=((\rho_{\rm rad}+\rho_\phi)/\rho_{\rm rad})^{3/4}$, at the
decay time of $\phi$, and multiply the primordial baryon-to-entropy
ratio by $D^{-1}$ to obtain the resulting baryon asymmetry.

\begin{figure}
\centerline{\epsfxsize=0.8\textwidth \epsfbox{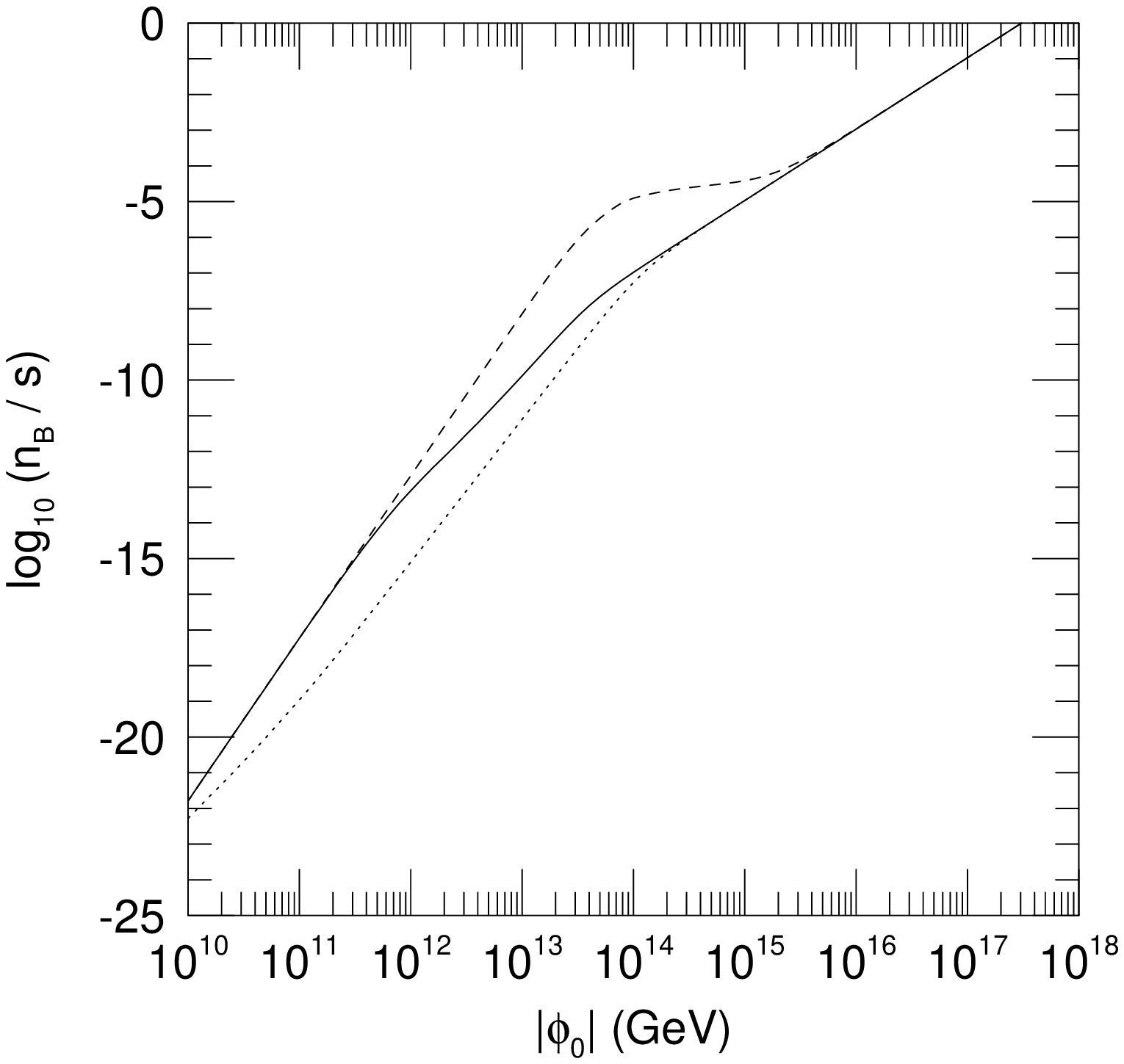}}
\caption{The resulting baryon-to-entropy ratio as a function of the
initial amplitude $\phi_0$. The parameters are taken to be
$V_0=(3\times 10^3{\rm
GeV})^4$, $\theta_0=\pi/8$, $\phi_{\rm dec}=10^5$ GeV, 
and $m_{3/2}=1$ keV (dotted line),
$m_{3/2}=100$ keV (solid line), and $m_{3/2}=10$ MeV (dashed line).}
\label{fig:nb/s}
\end{figure}

In Fig.~\ref{fig:nb/s}, we show the $|\phi_0|$ dependence of the
present baryon-to-entropy ratio, $n_B/s$. From the figure, we can see
that the results based on the order of magnitude estimations provide us good
approximations. For a sufficiently large $\phi_0$ such that the entropy
production is significant, the resultant baryon-to-entropy ratio is
independent of the gravitino mass, and is proportional to
$|\phi_0|^2$. We also checked that the approximate formula
(\ref{nb/s/D}) reproduces the behavior for the large $|\phi_0|$ region.
For a smaller value of $|\phi_0| \lesssim 10^{14}$~GeV, the 
entropy production from the decay becomes negligible.  The result then 
is proportional to $|\phi_{0}|^{4}$ (Eq.~(\ref{nb/s})).  For an even
smaller $|\phi_{0}| \lesssim \phi_{\rm eq}\sim 7 \times 10^{11}~\mbox{GeV} 
\times (100~\mbox{keV}/m_{3/2})$, the behavior goes over to 
$|\phi_{0}|^{9/2}$ (Eq.~(\ref{nb/s-small})).  As noted in the paragraph 
below Eq.~(\ref{nb/s-small}), there is a transition region from $\sim 
10\phi_{\rm eq}$ to $\phi_{\rm eq}$, where the curves fall less 
steeply because a logarithmic enhancement factor comes in.  In any 
case, the baryon-to-entropy can clearly be sufficiently large in this 
scenario, as required by the standard big-bang nucleosynthesis 
$n_{B}/s \sim 10^{-10}$, if the initial amplitude is larger than 
$10^{13-14}$ GeV.

A more precise estimate of the baryon asymmetry requires the
specification of the flat direction, the relevant operator, and the 
size of the initial amplitude.  The
usual caveat concerning the $B-L$ invariance applies:  If one employs
an operator which preserves $B-L$ symmetry like that generated by
SU(5) grand unified models, it may be wiped out again by the
electroweak sphaleron effect \cite{KRS}.  One needs to either preserve
the $B$ and $L$ asymmetries using a Bose condensate \cite{DMO}, or by
generating a $B-L$ asymmetry \cite{CDO,MY}.  In view of the
discussions of the next section, we find the protection of $B$ via a
Bose condensate is a likely scenario.

\subsection{Diluting Gravitinos}

In the LEGM models, the mass of the gravitino is about 100
keV, and its mass density exceeds the closure limit if $T_{\rm max}$
is larger than (1 -- 10) TeV, as discussed in Section~\ref{sec:MMY}. 
Since the reheating after the primordial inflation raises the
temperature typically above $T_{\rm RH} \gtrsim 10^8$~GeV or so, 
we assume that the
gravitinos were once thermalized, which is of course the worse case
scenario.  We will discuss whether the decay of the flat direction can
generate a large enough entropy to dilute gravitinos below the closure
limit. 

Before discussing the implication of the entropy production to the
gravitino, it is useful to estimate the freeze-out temperature of the
gravitino, $T_{\rm freeze}$, which is the temperature at which the
expansion rate of the Universe becomes comprable to the production rate of
the gravitino: $H(T_{\rm freeze})\sim \Sigma_{\rm tot}n_{\rm
rad}(T_{\rm freeze})$. By using the gravitino production cross section
given in Eq.~(\ref{Sigma_tot}), we obtain
 \begin{eqnarray}
T_{\rm freeze} \sim 200{\rm TeV}\times
\left(\frac{m_{3/2}}{100{\rm keV}}\right)^{2}
\left(\frac{m_{\rm G3}}{1{\rm TeV}}\right)^{-2}.
\label{T_freeze}
 \end{eqnarray}
Below $T_{\rm freeze}$, the expansion rate of the Universe
becomes larger than the production rate of the gravitino, and hence the
gravitino cannot be thermalized. Therefore, 
an entropy production at $T\lesssim T_{\rm freeze}$ dilutes
gravitinos produced before the entropy production. On
the contrary, even if the entropy is produced when $T\gtrsim T_{\rm
freeze}$, the gravitino is thermalized again, and its number density
is determined by the thermal distributions. If the energy density of
the flat direction dominates the energy density of the Universe, decay
of the flat direction $\phi$ reheats the Universe. The reheating
temperature is estimated as $T_{\rm R}\sim (V_0)^{1/4}\sim 1-10$ TeV.
Comparing this reheating temperature to Eq.~(\ref{T_freeze}), we can
see that the gravitino cannot be thermalized after the decay of
$\phi$. In other words, gravitinos produced before the decay of $\phi$
are diluted with a dilution factor given in Eq.~(\ref{dilution-factor}),
if the decay of $\phi$ produces the entropy.

If the gravitino mass is larger than $\sim$ 1 keV, we need a substantial
entropy production; otherwise, the Universe is overclosed by the mass
density of the gravitino. By assuming that the gravitino is
thermalized, we can estimate the number density of the gravitino as
 \begin{eqnarray}
n_{3/2}(T) = \frac{3}{2}\frac{g_*(T)}{g_*(T_{\rm freeze})}
n_{\rm rad} (T).
 \end{eqnarray}
Then, requiring $\Omega_{3/2}=D^{-1}m_{3/2}n_{3/2}/\rho_{\rm c} \leq
1$, we obtain
 \begin{eqnarray}
 D\gtrsim 50\times h^{-2} \left(\frac{m_{3/2}}{\rm 100keV}\right).
 \label{D>50}
 \end{eqnarray}
Comparing the above constraint with Eq.~(\ref{dilution-factor}), we
can see that the decay of the flat direction can produce enough
entropy to dilute the gravitinos away. For example, for $\phi_{\rm
dec}\sim 10^5$ GeV and $m_{3/2}\sim$ 100 keV, the dilution factor is
large enough, if $|\phi_0|\gtrsim 10^{14-15}$ GeV. Even with such a
large dilution,
the present baryon-to-entropy ratio can be sufficiently large (see
Eq.~(\ref{nb/s/D})).

If the flat direction decays at an amplitude larger than $\sim 10^5$
GeV, the dilution factor given in Eq.~(\ref{dilution-factor}) becomes
smaller, and the entropy production due to the decay of $\phi$ may not
be enough to decrease the gravitino density. In addition, for
$m_{3/2}\lesssim 100$ keV, the reheating temperature $T_{\rm R}\sim
V_0^{-1/4}\sim 1-10$ TeV may be higher than the freeze-out temperature
of the gravitino. In these cases, we have to assume an extra source of
the entropy production of $O(m_{3/2}/2h^2{\rm keV})$ after the
reheating. 
Even in this case, the estimation of the primordial 
baryon-to-entropy ratio (\ref{nb/s}) is still valid, and the final 
baryon asymmetry can be as large as the one estimated by
Eq.~(\ref{nb/s/D}) {\it and} an 
additional dilution factor Eq.~(\ref{D>50}) required to dilute the 
gravitinos.  Therefore, Affleck-Dine baryogenesis can generate a big 
enough baryon asymmetry to explain the present value of the 
baryon-to-entropy ratio.

In fact, we can crudely estimate the decay amplitude $\phi_{\rm dec}$
even when it is in the region of the logarithmic potential, if its
motion is circular rather than elliptic.  The change from the original
Affleck--Dine estimate of the decay rate Eq.~(\ref{Gammaphi}) is that
the rotation frequency of the $\phi$ field is given by $(V_0 \ln
(|\phi|^2/\langle S \rangle^2))^{1/2}/|\phi|$ rather than $m_{3/2}$.  
Since we are interested in a dilution factor, we assume that the flat
direction dominates the Universe, and the field decays when
\begin{equation}
\Gamma_\phi \sim \left( \frac{\alpha_s}{\pi} \right)^2
	\frac{1}{|\phi_{\rm dec}|^2} 
	\left( \frac{V_0 \ln (|\phi_{\rm dec}|^2/\langle S \rangle^2)}
		{|\phi_{\rm dec}|^2} \right)^{3/2}
	\sim H \sim \frac{V_0^{1/2}}{M_*} \ln 
	\frac{|\phi_{\rm dec}|^2}{\langle S \rangle^2} .
\end{equation}
and
\begin{equation}
|\phi_{\rm dec}| \sim \left[ \left(\frac{\alpha_s}{\pi}\right)^2
		V_0 M_* 
		\left(\ln \frac{|\phi_{\rm dec}|^2}{\langle S \rangle^2}
			\right)^{1/2} \right]^{1/5}
	\sim 8\times 10^5~\mbox{GeV} \times 
		\left( \frac{V_{0}}{(3\times 10^{3}~\mbox{GeV})^{4}}
			\right)^{1/5}.
\end{equation}
Therefore, the decay amplitude does not change much from the value
assumed before.  On the other hand, the case with an elliptic orbit is
more difficult to deal with.  We are not aware of any study on
the decay rate of $\phi$ for an arbitrary elliptic motion even for the
parabolic potential.  The other limit of almost linear motion is
discussed in the literature and tends to give a larger decay rate, and
hence a larger $\phi_{\rm dec}$ \cite{DK}.  However, we believe the
motion of the $\phi$ field in the case of our interest here, namely for
$|\phi_0| \gtrsim 10^{14-15}$~GeV to dilute gravitinos, to be
quite far from a linear one, and $\phi_{\rm dec}$ is
not likely to be much larger than our estimate.
We conclude that 
$\phi_{\rm dec}$ is not much larger than the minimum possible value 
$10^{5}$~GeV which we used in most of our discussions.

\section{Cosmology of String Moduli}
\label{sec:moduli}
\setcounter{equation}{0}
\setcounter{footnote}{0}

We point out in this section that the moduli fields in the string 
theory, if they acquire masses in the LEGM models, are stable and 
drastically overclose the Universe.

According to a general analysis \cite{Casas}, string moduli acquire
masses comparable to the gravitino mass $m_{3/2}$.
Their initial amplitude is likely to be of the order of the string
or Planck scale because it is the only scale in the problem.  The
cosmological problem of the moduli fields is discussed extensively in
the literature in the context of hidden sector models (for the original
paper, see \cite{CHRR}).  There, the moduli fields acquire masses of the
order of 1~TeV, and decay after nucleosynthesis, thereby spoiling the
success of the nucleosynthesis theory.  Even if one pushes the mass to
10~TeV so that the moduli fields decay before nucleosynthesis, the
enormous production of entropy with a dilution factor of order
$M_{*}/m_{3/2} \sim 10^{14}$ wipes out all pre-existing baryon
asymmetry. This problem may be solved by adopting the Affleck-Dine
baryogenesis \cite{MYY}, or by the thermal inflation \cite{LS} (see 
Appendix~\ref{app:hidden} for more discussions).

In the LEGM models, the situation is completely different.\footnote
{It was argued that the problem does not exist \cite{BKN} if SUSY
is broken dynamically, which is true for scalar fields which directly
participate in the dynamical supersymmetry breaking.  However, the
string moduli fields were not considered in this discussion.}
The string moduli are {\it stable}\/ within the cosmological time
scale, and are still oscillating around their potential minima.  A
dimensional analysis gives the decay rate of a moduli field to be
$\Gamma \sim m_{3/2}^{3}/8\pi M_{*}^{2} \sim
(3 \times 10^{18}~\mbox{years})^{-1}$, for $m_{3/2}\sim 100$ keV. 
Therefore
there is a problem concerning its energy density.

The estimation of the moduli energy density is straight-forward.  
When a moduli field begins to oscillate, the expansion rate is $H 
\sim m_{3/2}$.  The entropy at this stage is given by $s \sim 
g_*^{1/4} (m_{3/2} M_{*})^{3/2}$.  Assuming the initial amplitude of 
order $M_{*}$, the ratio of the moduli energy density to the entropy 
is given by
 \begin{equation}
 	\frac{\rho_{\rm moduli}}{s} \sim
 	 g_*^{-1/4} (m_{3/2} M_{*})^{1/2}
 	 \sim 1.3 \times  10^{6}~\mbox{GeV}
 			\left(\frac{m_{3/2}}{100~\mbox{keV}} \right)^{1/2} .
 			\label{rho-moduli}
 \end{equation}
Since both the energy density of the moduli and the entropy are 
diluted by the expansion with the same rate $R^{-3}$, the ratio 
remains constant until now unless there is entropy production.  On 
the other hand, the total energy density is bounded from above by the 
critical density $\rho_{c}$,
 \begin{equation}
	\frac{\rho_{\rm moduli}}{s} \leq \frac{\rho_{c}}{s_{\rm now}} = 
		3.6\times 10^{-9} h^{2}~\mbox{GeV}.
	\label{rho-crit}
 \end{equation}
 where $s_{\rm now}$ is the present value of the entropy density.
The predicted ratio Eq.~(\ref{rho-moduli}) is in gross conflict 
with the constraint Eq.~(\ref{rho-crit}).

It is not clear how such an enormous energy density can be diluted.  
First of all, the necessary dilution factor is at least $10^{15}$.  
Furthermore, one needs such an entropy production at a very late stage 
of the Universe, with $H \leq m_{3/2}$, or equivalently, $T \leq 
g_*^{-1/4} (m_{3/2} M_{*})^{1/2} = 1.3 \times 10^{6}~\mbox{GeV}$ for 
$m_{3/2} = 100$~keV.  One needs to create a baryon asymmetry either 
after such an enormous entropy production at a very late stage, or 
large enough to survive the enormous entropy production.

Actually, the Affleck--Dine mechanism may create a large enough baryon
asymmetry to survive the enormous entropy production which dilutes the
string moduli below the critical density as we will show below.

A quantity which remains constant over an entropy production is the
ratio of the baryon number and the moduli energy density, because both
of them scale as $R^{-3}$.  We estimated in the previous section that
the initial baryon number density is $n_{B} \sim m_{3/2} {\rm
Im}\phi_{0}^{4}/M_{*}^{2}$, at the time when the flat direction begins
to oscillate, {\it i.e.}\/, $H \sim m_{3/2}$.  In fact, this is the
same time as when the moduli fields begin to oscillate, and the energy
density of the moduli is $\rho_{\rm moduli} \sim m_{3/2}^{2}
M_{*}^{2}$.  Therefore, their ratio is determined at this stage:
 \begin{equation}
	\frac{\rho_{\rm moduli}}{n_{B}} \sim
		m_{3/2} \left({\rm Im}
		\frac{M_{*}^{4}}{\phi_{0}^{4}}\right) .
 \end{equation}
 or equivalently, 
 \begin{eqnarray}
	\frac{n_B}{s} \sim \frac{\rho_{\rm moduli}}{s}\times
	m_{3/2}^{-1} \left(\frac{|\phi_0|}{M_*}\right)^4 \sin 4\theta_0.
 \label{nb/s(with_Polonyi)}
 \end{eqnarray}
 Combining the above equation with the constraint (\ref{rho-crit}), we
find
 \begin{eqnarray}
	\frac{n_B}{s}\lesssim \frac{\rho_{c}}{s_{\rm now}}\times
	m_{3/2}^{-1} \left(\frac{|\phi_0|}{M_*}\right)^4 \sin 4\theta_0
	\sim 4\times 10^{-5} h^{2} 
	\left(\frac{m_{3/2}}{100~\mbox{keV}}\right)^{-1}
	\left(\frac{|\phi_0|}{M_*}\right)^4 \sin 4\theta_0.
 \end{eqnarray}
 As one can see, if $|\phi_0|\gtrsim 10^{17}$ GeV, the
baryon-to-entropy ratio may be larger than $\sim 10^{-10}$ even if we
assume a large entropy production to dilute the moduli field. 

The important question is whether one can have a brief period of 
inflation at such a late stage of Universe to dilute string moduli in 
the LEGM models.  The inflationary expansion rate $H_{\rm inf}$ must 
be less than $H_{\rm inf} \lesssim m_{3/2} \sim 100$~keV, {\it i.e.}\/ 
the energy density of the inflation $\rho_{\rm inf} \lesssim (m_{3/2} 
M_*)^2 \sim (10^7~\mbox{GeV})^2$.  Moreover, the $e$-folding should 
not exceed 20 or so in order to keep the primordial density 
fluctuation generated by a ``standard'' inflation with $H_{\rm inf} 
\sim 10^{11}$--$10^{13}$~GeV \cite{RT}. On the other hand, 
an $e$-folding of $N \gtrsim 5$ is sufficient to dilute the string 
moduli by $10^{-15}$. A thermal inflation \cite{LS} may offer a 
natural solution to these questions.

Fortunately, we apparently do not need to introduce new energy scales
into the model in the framework of the thermal inflation.
Suppose a positive mass squared of $m^2 \sim (100~\mbox{GeV})^2$ is
generated for a scalar field $\chi$ due to higher order loops at the
energy scale of $\Lambda_{\rm DSB} \sim 10^7$~GeV. The
renormalization group running of the mass squared may drive it
negative at a scale slightly below $\Lambda_{\rm DSB}$.  If the scalar
field is a flat direction of both $F$- and $D$-terms in the potential,
it develops a minimum at $v \lesssim \Lambda_{\rm DSB}$.  This is an
ideal potential for a thermal inflation.  The scalar field may
initially be
stuck at the origin because of the thermal effects, giving a
cosmological constant.  As the radiation gets red-shifted, the thermal
effects turn off and the field rolls down the potential to its true
minimum $\chi = v$.  The $e$-folding in this case is roughly $N \simeq
\frac{1}{2} \ln (v/m) \sim \frac{1}{2} \ln (\Lambda_{\rm DSB}/m) \sim
5$ \cite{LS} which is exactly what is needed to dilute the 
string moduli below the critical density.

It may be interesting to compare this result with the case of the
hidden sector SUSY breaking scenario with the Polonyi field or with
the string moduli. (Hereafter, we call them generically as ``Polonyi fields''.   
See 
Appendix~\ref{app:hidden} for the estimation of baryon asymmetry in 
this case.)  Note that Eq. (\ref{nb/s(with_Polonyi)}) is valid. Thus, the
question is the constraint on the energy density of the Polonyi field,
$\rho_z$.  In this case, the Polonyi field decays much faster than the
LEGM case since its mass is larger, and a typical lifetime for the
Polonyi field is given by
 \begin{eqnarray}
	\tau_z \sim \left(\frac{N_{\rm ch}}{4\pi} 
	\frac{m_{3/2}^3}{M_*^2}\right)^{-1} \sim
	10^{3}{\rm sec}\times \left(\frac{m_{3/2}}{\rm 1TeV}\right)^3,
 \end{eqnarray}
 where $N_{\rm ch}\sim O(10)$ is the number of the decay channel.
Thus, it does not contribute to the mass density of the present
Universe, and the constraint (\ref{rho-crit}) cannot be applied.
However, it may affect the great success of the standard big-bang
nucleosynthesis (BBN) scenario. 

The mass density of the
Polonyi field speeds up the expansion rate of the Universe when the
neutron decouples from the thermal bath ({\it i.e.}, $T\sim$ 1MeV),
which may result in an over production of $^4$He. Furthermore, the
radiative decay of the Polonyi field induces cascade photons which
cause the photofission process and change the primordial abundances of
the light nuclei.  The constraint on the primordial density of the
Polonyi field strongly depends on its lifetime $\tau_z$
\cite{KM}. If $\tau_z\lesssim 10^4$ sec, 
nucleosynthesis requires $\rho_z/s\lesssim 10^{-5}$~GeV.
For a Polonyi field with a longer lifetime, the constraint becomes
more stringent.  For a Polonyi field with $\tau_z \lesssim 10^{(4-5)}$ 
sec, which is the case for
the Polonyi mass typically larger than a few TeV, its mass density is
constrained as $\rho_z/s\lesssim 10^{-7}$~GeV.  
These constraints on $\rho_{z}/s$ are compared to the 
estimate of the baryon-to-entropy ratio (\ref{nb-rhoz}) which holds 
irrespective of the 
presence of a substantial dilution of Polonyi field by, {\it e.g.}\/, 
a late inflation,
 \begin{displaymath}
	\frac{n_B}{s} \sim \frac{\rho_z}{s}\times
	m_{3/2}^{-1} \left(\frac{|\phi_0|}{M_*}\right)^4 \sin 4\theta_0.
 \end{displaymath}
Thus, for this range of the Polonyi
mass, the resulting
baryon-to-entropy ratio may still be as large as $10^{-10}$ if $|\phi_0|\sim
M_*$, and hence the Affleck-Dine scenario may provide us a reasonable
value for the baryon-to-entropy ratio.  However, if the Polonyi field
has a longer lifetime, as for a sub-TeV Polonyi mass as usually
expected, the constraint
on $\rho_z$ becomes even more stringent. In particular, for the case with
$\tau_z\gtrsim 10^7$ sec, which typically corresponds to $m_z\lesssim
100$~GeV, $\rho_z/s\lesssim 10^{-13}$~GeV. In this case, the result is too
small to be identified as the present baryon asymmetry of the
Universe.

\section{Conclusion}
\label{sec:conclusion}
\setcounter{equation}{0}
\setcounter{footnote}{0}

We studied the cosmology of the LEGM models.
We first estimated the lower bound on the gravitino mass, and saw 
that the
bound conflicts with the cosmological constraint if the primordial
gravitino is not diluted. This fact indicates a huge entropy
production at a relatively low temperature, and the conventional
scenario of baryogenesis may not work well. 

In this case, the Affleck-Dine baryogenesis is one interesting
possibility.
The size of the baryon number violating operators is much smaller
than in the hidden sector models.  However the flat direction begins to
move at a much later stage which in turn increases the baryon number.
The dilution factor due to the decay of flat direction also has a
complicated dependence on parameters.  
After putting all the effects together, we found that the Affleck--Dine
baryogenesis works efficiently for an
initial amplitude of the flat direction, $|\phi_{0}| \gtrsim 
10^{13}$~GeV.  We also discussed
that the decay of the MSSM flat direction may provide enough
entropy to dilute the primordial gravitino for a relatively large
initial amplitude of the flat direction, $|\phi_0| \gtrsim
10^{14-15}$~GeV.  Therefore,
the gravitino problem in the LEGM models may be solved if we assume such
a large initial amplitude.

We also discussed the cosmological implication of the moduli fields in
the string theory. Their masses are of the order of the gravitino
mass, and their lifetime is much larger than the present age of the
Universe in the LEGM models.  The mass density of the
moduli field may overclose the Universe. To
dilute the moduli fields, a very late inflation is needed. We found
that the baryon asymmetry generated by Affleck-Dine baryogenesis can be
large enough to survive such a late inflation for $|\phi_{0}| \gtrsim 
10^{17}$~GeV, even if we assume a huge entropy production to dilute 
the primordial moduli field below the critical density.

\section*{Acknowledgment}

This work was supported in part by the Director, Office of Energy
Research, Office of High Energy and Nuclear Physics, Division of High
Energy Physics of the U.S. Department of Energy under Contract
DE-AC03-76SF00098 and in part by the National Science Foundation under
grant PHY-95-14797.  HM was also supported by the Alfred P. Sloan
Foundation.  AdG was supported by CNPq (Brazil).

\appendix

\section{Computing the Effective Potential}
\label{app:potential}
\setcounter{equation}{0}
\setcounter{footnote}{0}


The effective potential for the flat direction $\phi$ can be computed 
by the following usual procedure.  We let it to have an expectation 
value, and calculate the vacuum energy in the presence of $\phi$ 
background.  The vacuum energy is identically zero if we do not pick 
the effect of supersymmetry breaking.  The lowest order contribution 
is from two-loop diagrams where the standard model gauge multiplets 
couple to the vector-like messenger fields whose mass spectrum breaks 
supersymmetry.  The gauge multiplets acquire masses because of the 
$\phi$ background and hence the result depends on $\phi$.

The mass spectrum of the messenger sector is $M$ for fermions, and 
$M_{\pm}^{2} = M^{2}\pm MB$ for scalars.  We assume one vector-like 
multiplet with a unit U(1) charge, and calculate the contribution from 
a U(1) gauge multiplet exchange.  This U(1) gauge group is the 
toy-model version of the standard model gauge groups.  The flat 
direction $\phi$, $\bar{\phi}$ also has $\pm 1$ charge under U(1), 
with $D$-flatness condition $\phi = \bar{\phi}$.  The result can be 
easily generalized to arbitrary gauge groups and messenger multiplets.  
We refer to the U(1) gauge coupling constant as $g$.  The U(1) gauge 
multiplet acquires a mass $m = 2 g |\phi|$.  The task is to calculate 
the vacuum energy as a function of $M$, $B$, $g$ and $m$.  

\begin{figure}
\centerline{\epsfxsize=0.4\textwidth \epsfbox{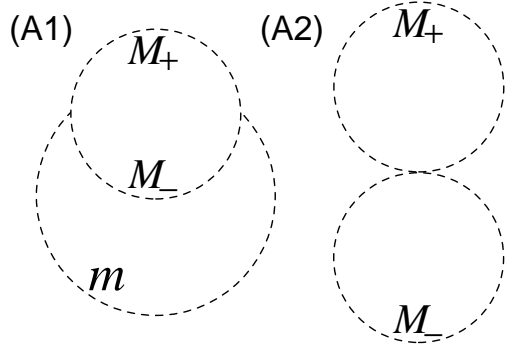}}
\caption[diagramA]{Feynman diagrams which contribute to the vacuum 
energy in the background of the flat direction $\phi = \bar{\phi}$.  
The vertices are due to the $D$-term potential.  The scalar field with 
mass $m=2g\langle \phi\rangle$ is the scalar component of the massive 
gauge multiplet in the presence of the background $\phi$.  The scalar 
fields with masses $M_{+}$ and $M_{-}$ are the messenger scalars.}
\label{fig:diagramA}
\end{figure}

\begin{figure}
\centerline{\epsfxsize=0.2\textwidth \epsfbox{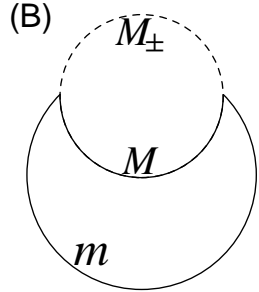}}
\caption[diagramB]{A Feynman diagram with the gaugino of mass $m$, 
the messenger fermion of mass $M$, and the messenger scalars of 
mass $M_{\pm}$.  }
\label{fig:diagramB}
\end{figure}

\begin{figure}
\centerline{\epsfxsize=0.6\textwidth \epsfbox{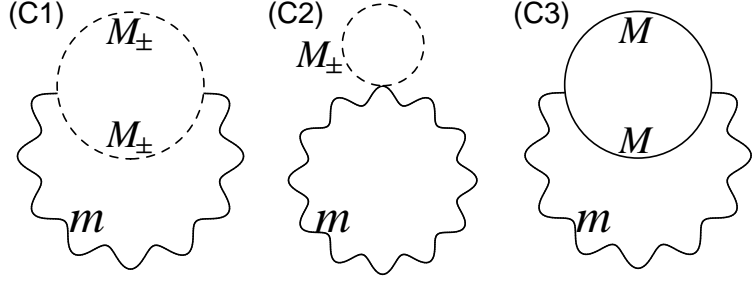}}
\caption[diagramC]{Feynman diagrams with vacuum polarization due to 
(C1) messenger scalar loops, (C2) ``seagull'' diagram with messenger 
scalars, and (C3) messenger fermions.}
\label{fig:diagramC}
\end{figure}

The Feynman
diagrams are shown in Figs.~\ref{fig:diagramA}--\ref{fig:diagramC}.
In all our calculations we expand the
amplitudes in terms of $MB/M^{2}$ and keep only the leading 
non-trivial terms of $O(MB)^{2}$.  We need $MB/M^{2} < 1$ to avoid a
color- or charge-breaking vacuum, and this expansion is known to be a 
good approximation for the supersymmetry breaking mass squared for the 
flat direction unless $MB$ is very close to $M^{2}$ \cite{private}.

We start with diagrams (A1) and (A2) in Fig.~\ref{fig:diagramA},
 \begin{eqnarray}
\mbox{(A1)}&=&ig^{2}m^{2}\int{d^{4}p\over{(2\pi)^4}}
\int{d^{4}k\over{(2\pi)^4}}{1\over{p^{2}-m^{2}}}\left[ 
{1\over{(k^{2}-M_{+}^{2})((p+k)^2-M_{-}^2)}} \right], \\ 
\mbox{(A2)}&=&ig^{2}\int{d^{4}p\over{(2\pi)^4}}
\int{d^{4}k\over{(2\pi)^4}}\left[ 
{1\over{(k^{2}-M_{+}^{2})(p^2-M_{-}^2)}} \right].  
 \end{eqnarray}

Since we know that the sum of all diagrams vanishes in the 
supersymmetric limit $MB \rightarrow 0$, we subtract the corresponding 
amplitude in the supersymmetric limit from each diagrams.  The 
diagrams (A1) and (A2) give after the subtraction:
 \begin{eqnarray}
\mbox{(A1)}_s +\mbox{(A2)}_s  &=&  
-ig^2\int_{0}^{1}dx\int{d^{4}p\over{(2\pi)^4}}
\int{d^{4}k\over{(2\pi)^4}}{p^2\over{p^2-m^2}} 
\nonumber \\
& & \left[ 
\frac{1}{(k^2+x(1-x)p^2-M^2+(1-2x)M\!B)^2}
- (MB \rightarrow 0)
\right]. 
 \end{eqnarray}
Here and hereafter, the subscript ${s}$ refers to the subtraction 
of amplitudes in the supersymmetric limit.

We expand the integrand in powers of $M\!B/M^2$.  The linear terms in 
$M\!B$ vanish upon $x$ integration, and we are left with the 
following expression, to $O((M\!B)^2)$ in the 
integrand,
 \begin{equation}
\mbox{(A1)}_{s}+\mbox{(A2)}_s =  
ig^2\int_{0}^{1}dx\int{d^{4}p\over{(2\pi)^4}}
\int{d^{4}k\over{(2\pi)^4}}{p^2\over{p^2-m^2}} 
{(4-1)(1-2x)^2(M\!B)^2\over{(k^2+x(1-x)p^2-M^2)^4}} + 
O(B^{4}).
 \end{equation}

We follow the same strategy as above to compute the contribution from 
the diagram (B) in Fig.~\ref{fig:diagramB} containing the messenger 
fermions,
 \begin{eqnarray}
\mbox{(B)}_s &=&  
-i (\sqrt{2} g)^{2} \int{d^{4}p\over{(2\pi)^4}}
\int{d^{4}k\over{(2\pi)^4}}
\frac{2 (k\cdot p)}{(p^{2} - m^{2})(k^{2} - M^{2})}
	\nonumber \\
& &
	\left( \frac{1}{(p+k)^{2} - M_{+}^{2}} 
		+ (M_{+}^{2} \rightarrow M_{-}^{2})
		- 2 (M_{+}^{2} \rightarrow M^{2}) \right)
		\nonumber \\
&=& -4ig^2\int_{0}^{1}dx\int{d^{4}p\over{(2\pi)^4}}
\int{d^{4}k\over{(2\pi)^4}}{xp^2\over{p^2-m^2}} 
\nonumber \\
& & \left[ 
{1\over{(k^2+x(1-x)p^2-M^2-xM\!B)^2}} + (MB \rightarrow - MB)
- 2 (MB \rightarrow 0) \right] \nonumber \\ 
&=& -4ig^2\int_{0}^{1}dx\int{d^{4}p\over{(2\pi)^4}}
\int{d^{4}k\over{(2\pi)^4}}{p^2\over{p^2-m^2}} 
{(M\!B)^26x^3\over{(k^2+x(1-x)p^2-M^2)^4}}
+O(B^{4}). \nonumber \\
 \end{eqnarray}


Finally the diagrams (C1), (C2), (C3) with the gauge boson loop in 
Fig.~\ref{fig:diagramC}.  We first calculate the vacuum 
polarization diagrams of messenger fields.  Note that the 
contribution of messenger fermions (C3) is the same as the one in the 
supersymmetric limit, and hence cancels after the subtraction.  The 
scalar loop gives
 \begin{eqnarray}
	\mbox{(C1)}_{s} &=&  g^{2} \int \frac{d^{4}k}{(2\pi)^{4}}
		(2k+p)^{\mu} (2k+p)^{\nu} 
		\nonumber \\
	& & 
		\left[ \frac{1}{k^{2} - M_{+}^{2}} 
		\frac{1}{(k+p)^{2} - M_{+}^{2}}
		+ (M_{+}^{2} \rightarrow M_{-}^{2}) 
		- 2 (M_{+}^{2} \rightarrow M^{2}) \right],
 \end{eqnarray}
and the ``seagull'' diagram gives
 \begin{equation}
	\mbox{(C2)}_{s} = -2g^{2} g^{\mu\nu} 
	\int \frac{d^{4}k}{(2\pi)^{4}} \left[
	\frac{1}{k^{2} - M_{+}^{2}} + (M_{+}^{2} \rightarrow M_{-}^{2}) 
	- 2 (M_{+}^{2} \rightarrow M^{2}) \right] .
 \end{equation}
Their sum is
 \begin{equation}
	\mbox{(C1)}_{s} + \mbox{(C2)}_{s}
	= g^{2} (-g^{\mu\nu} p^{2} + p^{\mu} p^{\nu})
	\int \frac{d^{4}k}{(2\pi)^{4}}
	\int_{0}^{1} dz  
	\frac{6 (1-2z)^{2}(MB)^{2}}{(k^{2} + z(1-z) p^{2} - M^{2})^{4}}
	+ O(B^{4}) .
 \end{equation}
Now including the gauge boson loop, the total 
contribution of the vacuum polarization diagrams is
 \begin{equation}
	\mbox{(C)}_{s} =
	i 9 g^{2} (MB)^{2}
	\int \frac{d^{4}k}{(2\pi)^{4}} \int \frac{d^{4}p}{(2\pi)^{4}}
	\int_{0}^{1} dz \frac{p^{2}}{p^{2} - m^{2}}
	\frac{(1-2z)^{2}}{(k^{2} + z(1-z) p^{2} - M^{2})^{4}}.
 \end{equation}

Adding all diagrams, we obtain $-i V_{\it eff}(m^{2}) = \mbox{(A)}_{s} + 
\mbox{(B)}_{s} + \mbox{(C)}_{s}$ up to $O(MB)^{2}$,
 \begin{equation}
V_{\it eff}(m^2)= 
-12g^2(M\!B)^2\int_{0}^{1}dx\int{d^{4}p\over{(2\pi)^4}}\int{d^{4}k\over{(2\pi)^4}}{p^2\over{p^2-m^2}}\left[ 
{(1-2x)^2-2x^3\over{(k^2+x(1-x)p^2-M^2)^4}} \right].
 \end{equation}
After a Wick rotation in the $k$ and $p$ spaces, the $d^4k$ 
integration can be carried out trivially. 
Note also that the denominator is symmetric under the interchange of 
$x\leftrightarrow(1-x)$. We therefore symmetrize the $x$ integration 
by substituting the polynomial in the numerator $P(x) = (1-2x)^2-2x^3$ by 
${1\over2}(P(x)+P(1-x)) = -x(1-x)$, finding
 \begin{equation}
V_{\it eff}(m^2)={-g^2(M\!B)^2\over{128\pi^4}}\int_{0}^{1}dx\int_{0}^{\infty}{p^2d(p^2)}\left[ 
{x(1-x)p^2\over{(p^2+m^2)(x(1-x)p^2+M^2)^2}} \right].
 \end{equation}
The $p^{2}$ integral is logarithmically divergent.  Fortunately, the 
divergent piece is $m^{2}$ independent, and hence
is the renormalization of the cosmological constant.
We subtract
$V_{\it eff}(0)$ from the above expression and redefine it as $V_{\it 
eff}$. The final integral is convergent for any $m^2\in[0,\infty)$:
 \begin{equation}
V_{\it eff}(m^2)={g^2(M\!B)^2m^2\over{128\pi^4}}\int_{0}^{1}dx\int_{0}^{\infty}{p^2d(p^2)}\left[ 
{x(1-x)\over{(p^2+m^2)(x(1-x)p^2+M^2)^2}} \right].
 \end{equation}
The $p$ integration can be computed using the following tricks.
First, change the integration variable to $q^{2} = x(1-x) p^{2}$.
Then the $q^{2}$ integration can be done in the standard way, and we obtain 
 \begin{equation}
V_{\it eff}(z^2)
 =   {g^2(M\!B)^2\over{128\pi^4}}\int_{0}^{1}dx\left[ 
{1/z^2-x(1-x)+x(1-x)\ln[x(1-x)z^2]\over{(1/z^2-x(1-x))^2}} \right].
 \end{equation}
Here and below, we use $z^2\equiv {m^2/{M^2}}$.

\begin{figure}
\centerline{\epsfxsize=0.6\textwidth \epsfbox{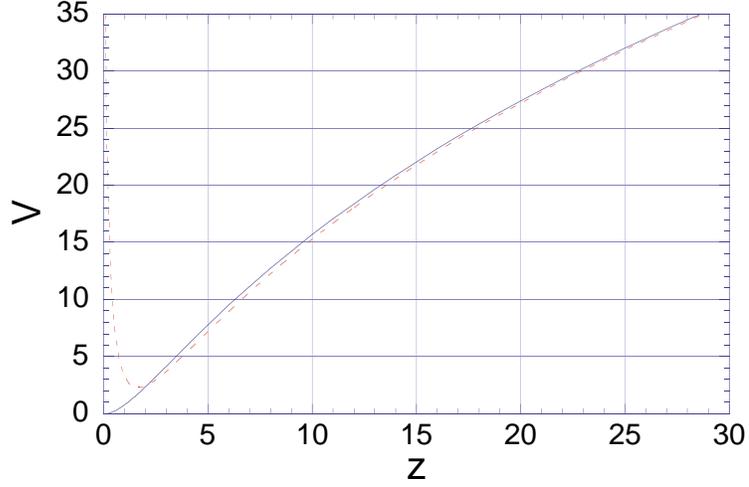}}
\caption[large-z]{A plot of the effective potential Eq.~(\ref{Veff}) 
for $z \equiv 2 g |\phi|/M< 30$, in the unit of 
$g^2(M\!B)^2/{(128\pi^4)}$.  The solid line is the exact result, and 
the dotted line shows the asymptotic form Eq.~(\ref{asymptotic}) valid 
for large $z$.} 
\label{fig:large-z}
\end{figure}

\newcommand{\Li}{\mbox{Li}}
We can further perform the $x$ integration using dilogarithms.  Using
the roots of the denominator $a\equiv \left(1 - \sqrt{1-4/z^2}\right)/2$ 
and $1-a$,
 \begin{equation}
V_{\it eff}(a)={g^2(M\!B)^2\over{128\pi^4}}\int_{0}^{1}dx \left[ 
{(x-a)(x-1+a)+x(1-x)\ln\left( {x(1-x)\over{a(1-a)}} \right) 
\over{(x-a)^2(x-1+a)^2}} \right].
 \end{equation}
After the final integral is carried out we are left with an
expression for the effective potential as a function of $a$:
 \begin{eqnarray}
V_{\it eff}(a)&=&
{g^2(M\!B)^2\over{64\pi^4}}\left\{{\ln(a(1-a))\over{(1-2a)^2}}+\right. 
\nonumber \\
&&\left.{1-2a(1-a)\over{(1-2a)^3}}
\left[{1\over{2}}\ln^2(a)-{1\over{2}}\ln^2(1-a)-\Li_{2}(a)+\Li_{2}(1-a)
	\right]\right\}.
	\label{Veff}
 \end{eqnarray}
The form of the effective potential is shown in Fig.~\ref{fig:large-z} as 
a function of $z$ as a solid line.  

The expression is manifestly real for $z^2\geq4$. In the limit 
$z\rightarrow\infty$, $a\approx1/z^2$ and the potential behaves as
 \begin{eqnarray}
V_{\it eff}
&=&{g^2(M\!B)^2\over{64\pi^4}}\left[{1\over{2}}\ln^2(z^2)-\ln(z^2)
+\frac{\pi^{2}}{6} + O\left(\frac{1}{z^{2}} \ln^{2} z^{2}\right)
\right]
\nonumber\\
&\sim & {g^2(M\!B)^2\over{128\pi^4}} 
\left( \ln \frac{4 g^{2} |\phi|^{2}}{e M^{2}} \right)^{2}.
\label{asymptotic}
 \end{eqnarray}
This asymptotic form of the effective potential is also shown in 
Fig.~\ref{fig:large-z} as a dotted line.

\begin{figure}
\centerline{\epsfxsize=0.6\textwidth \epsfbox{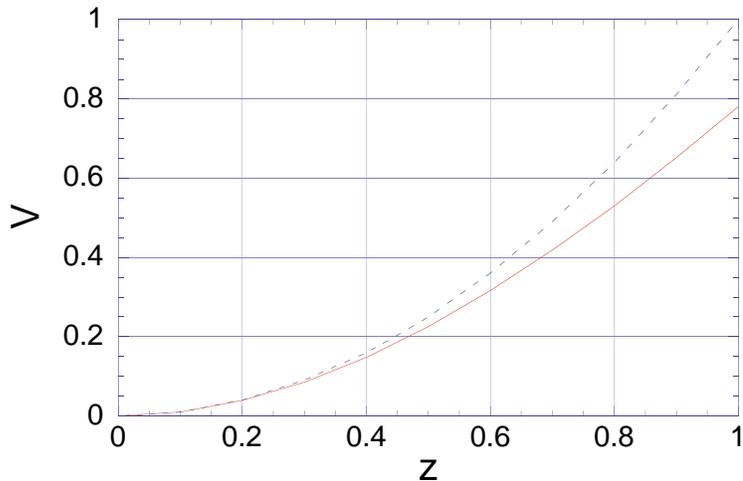}}
\caption[small-z]{A plot of the effective potential Eq.~(\ref{Veff})
(or equivalently, Eq.~(\ref{Veff-smallz}))
for the small field amplitude, $z \equiv 2 g |\phi|/M< 1$, in the unit 
of $g^2(M\!B)^2/{(128\pi^4)}$.  The solid line is the exact result, 
and the dotted line shows the approximate form Eq.~(\ref{small-z}) 
valid for small $z$.}
\label{fig:small-z}
\end{figure}

In the case $z^2<4$, $a$ is complex, $a=1/2+i\sqrt{4/z^2-1}$.  We can 
make the effective potential manifestly real, using the following 
dilog relations:
 \begin{eqnarray}
\Li_{2}(a)+{1\over{2}}\ln^2(a)&=&-\Li_{2}
\left({1\over{a}}\right)-i\pi\ln(a)+{\pi^2\over{2}}-\zeta(2), 
\\
\Li_{2}(1-a)+{1\over{2}}\ln^2(1-a)&=&
-\Li_{2}\left({1\over{1-a}}\right)+i\pi\ln(1-a)+{\pi^2\over{2}}-\zeta(2).
 \end{eqnarray}
We find
 \begin{eqnarray}
\lefteqn{
V_{\it eff}(z^2<4)={g^2(M\!B)^2\over{64\pi^4}}
\left\{-{\ln\left({1\over{z^2}}\right)\over{{4\over{z^2}}-1}}
\right.
}
\nonumber \\
& &
\left.
+{\left(1-{2\over{z^2}}\right)
\over{\left({4\over{z^2}}-1\right)^{3\over{2}}}}
\left[-\ln\left({1\over{z^2}}\right)
\left(\pi+\arctan\left(-\sqrt{{4\over{z^2}}-1}\right)\right)
+i\left(\Li_2\left({1\over{a}}\right)
-\Li_2\left({1\over{1-a}}
\right)\right)\right]\right\}.
\nonumber \\
\label{Veff-smallz}
 \end{eqnarray}
Note that $\Li_{2}(z)-\Li_{2}(z^*)$ is pure imaginary. In the above form, 
it is simple to take the limit $z^2\rightarrow0$, and we obtain
 \begin{eqnarray}
V_{\it eff}
&=&{g^2(M\!B)^2\over{64\pi^4}}
\left[{z^2\over{4}}\ln(z^2)+\left({z^3\over{8}}-{z\over{4}}\right)
\left(z\ln(z^2)+{iz^2\over{2}}-z-{iz^2\over{2}}-z\right)+O(z^3\ln(z^2))\right] 
\nonumber \\
&=&{g^2(M\!B)^2\over{64\pi^4}}\left({z^2\over{2}}+O(z^3\ln(z^2))\right) 
\label{small-z} \\
&=&{g^2m^2\over{128\pi^4}}\left({(M\!B)^2\over{M^2}}\right)+O(m^4)
\nonumber \\
&=& 4 \left( \frac{\alpha}{4\pi} \right)^{2}
	\left( \frac{M\!B}{M} \right)^{2}
	(|\phi|^{2} + |\bar{\phi}|^{2}) + O(\phi^{4}).
	\label{mass}
 \end{eqnarray}
The approximate form Eq.~(\ref{small-z}) truncated at $O(z^{2})$ is 
shown in Fig.~\ref{fig:small-z} as a dotted line together with the exact 
form Eq.~(\ref{Veff}) (or equivalently, Eq.~(\ref{Veff-smallz})) as a 
solid line.
From the last expression (\ref{mass}) with $\alpha \equiv g^{2}/4\pi$, we can read 
off the mass of the flat direction.  For messengers in ${\bf 5}+{\bf 
5}^{*}$ representation, we multiply the final result by a group 
theory factor $T^{a} T^{b} \mbox{tr}(T^{a} T^{b}) = \frac{1}{2} C_{f}$ 
where the trace is taken over the messenger fields and $C_{f}$ is the 
second order Casimir for the flat direction.  We obtain
$m_{\phi}^{2} = 2 C_{f}\left( \frac{\alpha}{4\pi} \right)^{2}
\left( \frac{M\!B}{M} \right)^{2}$, which agrees with 
that in Ref.~\cite{DN3}

\section{Effective Potential and Wave-function Renormalization}
\label{app:wavefn}
\setcounter{equation}{0}
\setcounter{footnote}{0}

When one computes an effective potential, one can determine the 
location of the minimum.  It is well-known that one also needs to 
evaluate the wave-function renormalization $Z(\phi) (\partial 
\phi)^{2}$ in order to discuss the time evolution of the scalar 
field in general.  Fortunately, such a calculation is not necessary in our case.

Let us recall the simple fact that the effective potential in our 
case is at 2-loop order: $V \sim (\alpha/4\pi)^{2}$.  Since it is 
a flat direction in the supersymmetric limit, this is the {\it 
only}\/ term in the potential.  The equation of motion in the flat 
space is
 \begin{equation}
	\ddot{\phi} + \frac{Z'(\phi)}{Z(\phi)} (\dot{\phi})^{2} + 
		\frac{1}{Z(\phi)} V'(\phi) = 0 .
 \end{equation}
Here we dropped the friction term $3H\dot{\phi}$, but the essence of the 
following discussions does not depend on such simplifying assumptions.

Because $V'$ is of order $(\alpha/4\pi)^{2}$, the motion is suppressed
by a power in the coupling constant.  Note that $Z'(\phi)$ is {\it at 
most}\/ order $(\alpha/4\pi)$.  By factoring out the coupling 
constant factors,
 \begin{eqnarray}
	V &=& \left(\frac{\alpha}{4\pi}\right)^{2} v(\phi) \\
	Z &=& 1 + \left(\frac{\alpha}{4\pi}\right) \zeta(\phi),
 \end{eqnarray}
we find
 \begin{equation}
	\ddot{\phi} + 
		\left(\frac{\alpha}{4\pi}\right) 
			\frac{\zeta'(\phi)}{Z(\phi)} (\dot{\phi})^{2}
		+ \frac{1}{Z(\phi)} 
			\left(\frac{\alpha}{4\pi}\right)^{2} v' (\phi) = 0.
 \end{equation}
It is convenient to rescale the time variable $t$ by 
 \begin{equation}
	\tau \equiv \frac{\alpha}{4\pi} t,
 \end{equation}
and we find
 \begin{equation}
	\frac{\partial^{2} \phi}{\partial \tau^{2}}
		+ \left(\frac{\alpha}{4\pi}\right) 
			\frac{\zeta'(\phi)}{Z(\phi)} 
			\left( \frac{\partial \phi}{\partial \tau} \right)^{2}
		+ \frac{1}{Z(\phi)} v'(\phi) = 0.
 \end{equation}
It is clear that the leading terms in the equation of motion are given 
by $\partial^{2} \phi/\partial \tau^{2} + v'(\phi) = 0$, and all 
dependences on the wave function renormalization occur only at higher 
orders in perturbation theory.  Therefore, the calculation of the 
effective potential is enough for our purpose, and we do not need 
$Z(\phi)$.

\section{Time Evolution of the Flat Direction}
\setcounter{equation}{0}
\label{app:virial}
\setcounter{footnote}{0}

The evolution of the flat direction $\phi$ is interesting in the LEGM
models. Once the amplitude is dominated by the gauge-mediated piece,
the potential is approximately proportional to $(\ln
|\phi|^2)^{2}$, and the dilution of the coherent oscillation occurs
much slower than in the parabolic potential case. In this Appendix, we
investigate the evolution of the flat direction by using the virial
theorem.

The virial theorem tells us that,
 \begin{equation}
	2 \langle K \rangle
		= \left\langle \frac{\partial V}{\partial \phi} \phi 
		+ \frac{\partial V}{\partial \phi^*} \phi^*
		\right\rangle,
 \end{equation}
 where $K =\dot{\phi}^*\dot{\phi}$ is the kinetic energy.  In our
case, $V \sim V_{0} (\ln|\phi|^{2}/\langle S\rangle^2)^{2}$ with
$\langle S\rangle\sim 3\times 10^{4}$~GeV, and
 \begin{equation}
	\langle K \rangle
		= \left\langle 
		\frac{2}{\ln |\phi|^{2}/\langle S\rangle^2} V 
		\right\rangle .
 \end{equation}
For $\ln |\phi|^2/\langle S \rangle^2 \gg 1$,
the energy density of the field is potential dominated.

The field equation is
 \begin{equation}
	\ddot{\phi} + 3 H \dot{\phi} 
		+ \frac{\partial V}{\partial \phi^*} = 0.
 \end{equation}
 Multiplying it by $\dot{\phi}^*$ and using the energy density ${\cal
E} = K + V$,
 \begin{equation}
	0=\dot{\cal E} + 6 H K
		\sim \dot{\cal E} 
		+ 6 H \frac{2}{\ln |\phi|^{2}/\langle S\rangle^2} 
		{\cal E} .
 \end{equation}
Since the energy density is dominated by the potential term, we can
write down the approximate evolution equation of the amplitude $|\phi|$,
 \begin{equation}
	\frac{d|\phi|}{dt} \sim - 3 H |\phi|,
 \end{equation}
 and hence
 \begin{equation}
	|\phi(t)|  R(t)^3 \sim \mbox{constant}
 \end{equation}
 This formula is valid when $\ln|\phi|^2/\langle S\rangle^2 \gg 1$.

\section{Estimation of the Primordial Baryon Asymmetry}
\label{app:linde}
\setcounter{equation}{0}
\setcounter{footnote}{0}

In this Appendix, we justify Linde's formula, Eq.~(\ref{nb=nbdot/H})
in our case.  In fact, the validity of the formula depends on the
nature of the operator ${\cal O}$ and the time evolution of the flat
direction $\phi$.  We clarify the reason why the formula is valid in the
cases of our interest.
We start from the equation of motion for the baryon number density,
Eq.~(\ref{nBeom}), 
 \begin{equation}
	\dot{n}_{B} + 3 H n_{B} = i\left(
		\frac{\partial {\cal O}}{\partial \phi} \phi
		- \frac{\partial {\cal O}}{\partial \phi^{*}}
                \phi^{*}\right) .
 \end{equation}
It is useful to rewrite the equation in terms of baryon-to-entropy ratio
$Y_B \equiv n_B/s$, to find
\begin{equation}
\dot{Y}_B = \frac{1}{s} i\left(
		\frac{\partial {\cal O}}{\partial \phi} \phi
		- \frac{\partial {\cal O}}{\partial \phi^{*}}
                \phi^{*}\right) ,
\end{equation}
where we used the relation $sR^3=$ constant.
Assuming a vanishing initial value $Y_B (t_0) = 0$, we obtain
\begin{equation}
Y_B (\infty) = \int_{t_0}^\infty dt
		\frac{1}{s} i\left(
		\frac{\partial {\cal O}}{\partial \phi} \phi
		- \frac{\partial {\cal O}}{\partial \phi^{*}}
                \phi^{*}\right) .
\label{YBeom}
\end{equation}
A crucial question is whether the $t$ integral is dominated at $t\sim
t_0$ or $t \sim \infty$.  

In the following analysis, we assume the Universe to be radiation dominated
when the field begins to roll down the potential, $R \propto t^{1/2}$.
Another assumption is that the baryon-number violating operator ${\cal
O}$ can be treated as a small perturbation to the evolution of the field
$\phi$.

In the case $\phi_0 \gtrsim \phi_{\rm eq}$, or in the general hidden sector
case, we have ${\cal O} \propto \phi^4$.  The evolution of $\phi$ is
essentially determined by $m_{3/2}^2 \phi^2$ by assumption and hence
$\phi \propto R^{-3/2} \propto t^{-3/4}$.  On the other hand, $s \propto
R^{-3} \propto t^{-3/2}$.  The integrand in Eq.~(\ref{YBeom}) therefore
behaves as $t^{-3/2}$ and hence it is dominated by $t\sim t_0$.  By
putting them together,
\begin{eqnarray}
Y_B (\infty) &=& \int_{t_0}^\infty dt
		\frac{1}{s_0} i\left(
		\frac{\partial {\cal O}}{\partial \phi} \phi
		- \frac{\partial {\cal O}}{\partial \phi^{*}}
                \phi^{*}\right)
		\frac{t^{-3/2}}{t_0^{-3/2}}
	\nonumber \\
	&=& \frac{1}{2} t_0\times  
	\frac{8 m_{3/2}^2\mbox{Im}(\phi_0^4)}{s_0M_*^2} 
	= \frac{2m_{3/2}^2\mbox{Im}(\phi_0^4)}{s_0M_*^2H_0},
\end{eqnarray}
which essentially justifies Eq.~(\ref{nb=nbdot/H}).

In the other case of interest, $\phi_0 \lesssim \phi_{\rm eq}$, both
the behavior of the operator and time-evolution are completely different
as discussed in detail in Section~\ref{sec:ADinLEGM}.  We have ${\cal
O}\propto \phi^2$, while $\phi \propto R^{-3} \propto t^{-3/2}$.  Then
the integrand behaves as $t^{-3/2}$, which is unexpectedly the same as
in the previous case.  By putting them together, we obtain
\begin{eqnarray}
Y_B (\infty) &=& \int_{t_0}^\infty dt
		\frac{1}{s_0} i\left(
		\frac{\partial {\cal O}}{\partial \phi} \phi
		- \frac{\partial {\cal O}}{\partial \phi^{*}}
                \phi^{*}\right)
		\frac{t^{-3/2}}{t_0^{-3/2}}
	\nonumber \\
	&=& \frac{1}{2} t_0\times  
	\frac{8 V_0\mbox{Im}(\phi_0^4)}{s_0|\phi_0|^2M_*^2} 
	= \frac{2V_0\mbox{Im}(\phi_0^4)}{s_0|\phi_0|^2M_*^2H_0},
\end{eqnarray}
which again essentially justifies Eq.~(\ref{nb=nbdot/H}).

As it is clear from above the derivations, 
Eq.~(\ref{nb=nbdot/H}) is not
necessarily valid if the integral is dominated at $t\sim \infty$ rather
than $t\sim t_0$.  We have not seen an explicit discussion on this point
in the literature.

\section{Affleck-Dine Baryogenesis in Hidden Sector Scenario}
\label{app:hidden}
\setcounter{equation}{0}
\setcounter{footnote}{0}

In this appendix, we discuss the Affleck-Dine baryogenesis based on
the scenario with SUSY breaking in the hidden sector. In this case,
the gravitino mass is much larger than in the LEGM case, and all the scalar
fields also have the SUSY breaking masses of the order of the gravitino
mass.\footnote
 {In this Appendix, we denote all the soft SUSY breaking masses for
the scalar fields by $m_{3/2}$ for simplicity.}
 In particular, the potential for the flat direction is always
given by the supergravity contribution, which is essentially
parabolic with a curvature of the order of the gravitino mass,
 \begin{eqnarray}
	V(\phi)\sim m_{3/2}^2 |\phi|^2, 
  \end{eqnarray}
 with $m_{3/2}\sim$ 1 TeV. Due to this fact, the evolution of the flat
direction is much simpler than in the LEGM case.

Even if the gravitino mass is about 1 TeV, Eq.(\ref{nb=nbdot/H}) is
still valid since the baryon number is generated when $\phi$ starts to
oscillate. With the baryon number violating operator
(\ref{O_Bviolate}), we obtain
 \begin{eqnarray}
	n_B|_{H\sim m_{3/2}} \sim 
	\frac{m_{3/2} |\phi_0|^4}{M_{*}^{2}}\sin 4\theta_0.
 \label{nB(app)}
 \end{eqnarray}
 and hence
  \begin{eqnarray}
	\left. \frac{n_{B}}{s} \right|_{H\sim m_{3/2}}
	\sim g_*^{-1/4}
	\frac{|\phi_{0}|^{4}}{m_{3/2}^{1/2} M_{*}^{7/2}}
        \sin 4\theta_0.
 \label{nB/s(app)}
 \end{eqnarray}
 If there is no entropy production after this stage, the above
formula gives us the resulting baryon-to-entropy ratio.

If there is entropy production, the primordial baryon number is
diluted. The primary source of the entropy is the decay of the flat
direction. Here, the potential for the flat direction is always
parabolic, and $\phi$ starts to oscillate when $T=T_0\sim
g_*^{-1/4}\sqrt{m_{3/2}M_*}$, as discussed in Section~\ref{sec:ADinLEGM}.
Then, by using the relation $|\phi|^2T^{-3}$ = const., the background
temperature at the $\phi$ decay is given by
  \begin{eqnarray}
	T_{\rm dec} \sim g_*^{-1/4}\sqrt{m_{3/2}M_*} 
	\left(\frac{\phi_{\rm dec}}{|\phi_{0}|}\right)^{2/3},
 \label{T_dec(AD)}
 \end{eqnarray}
 where $\phi_{\rm dec}$ is the amplitude of the flat direction when it
decays. Furthermore, the reheating temperature due to the decay of
$\phi$, $T_{\rm R}$, is given by
  \begin{eqnarray}
	T_{\rm R} \sim g_*^{-1/4} \sqrt{m_{3/2}\phi_{\rm dec}}.
 \label{T_R(AD)}
 \end{eqnarray}
 Then, the dilution factor is given by
  \begin{eqnarray}
	D \sim \frac{T_{\rm R}^3}{T_{\rm dec}^3}
	\sim \frac{|\phi_0|^2}{\phi_{\rm dec}^{1/2}M_*^{3/2}}.
 \label{D(app)}
 \end{eqnarray}
 Usually, $\phi$ decays when the expansion rate of the Universe, $H$,
becomes comparable to the decay rate of $\phi$, $\Gamma_\phi$. In
Ref.\cite{AD}, $\Gamma_\phi$ is estimated as
  \begin{eqnarray}
	\Gamma_\phi \sim \left(\frac{\alpha_s}{\pi}\right)^2
	\frac{m_{3/2}^3}{|\phi|^2},
	\label{Gammaphi}
 \end{eqnarray}
 and hence $H\sim\Gamma_\phi$ results in
  \begin{eqnarray}
	\phi_{\rm dec} \sim \left(\frac{\alpha_s}{\pi}\right)^{2/3}
	m_{3/2}^{2/3} M_*^{1/3}.
 \end{eqnarray}
 Combining the above results, we obtain
  \begin{equation}
	\frac{n_B}{s} \sim
	\left(\frac{\alpha_s}{\pi}\right)^{1/3}
	\left(\frac{M_*}{m_{3/2}}\right)^{1/6}
	\left(\frac{|\phi_0|}{M_*}\right)^2
	\sin 4\theta_0
	= 120 \times \left(\frac{|\phi_0|}{M_*}\right)^2
	\left(\frac{\alpha_s}{0.1}\right)^{1/3}
	\left(\frac{m_{3/2}}{1~\mbox{TeV}}\right)^{-1/6}
	\sin 4\theta_0 .
	\label{nB/s-hidden}
 \end{equation}

Another potential source of entropy is the Polonyi field related
to the SUSY breaking, or the moduli fields in the string theory, which
also have masses of order $m_{3/2}$. The critical difference between
the flat direction and the Polonyi field $z$ is the formula of their decay
width; since the Polonyi field couples to particles in the observable sector 
only through interactions suppressed by the gravitational scale,
its decay width $\Gamma_z$ is much smaller than the width of $\phi$.
As discussed in Section~\ref{sec:moduli}, $\Gamma_z$ is estimated as
  \begin{eqnarray}
	\Gamma_z \sim \frac{N_{\rm ch}}{4\pi}\frac{m_{3/2}^3}{M_*^2}.
 \label{Gamma_z}
 \end{eqnarray}
 Even with this decay rate, we can apply an argument similar to the
case of the entropy production due to $\phi$ decay; Eqs.(\ref{nB(app)})
and (\ref{nB/s(app)}) are still valid, and we also obtain
equations similar to Eqs.(\ref{T_dec(AD)}) -- (\ref{D(app)}) where
$\phi$'s are replace by $z$'s. The remainder is to evaluate the
amplitude of $z$
at its decay time, $z_{\rm dec}$, by using the relevant formula for
$\Gamma_z$. By solving the equation $H\sim\Gamma_z$ with $H\sim
m_{3/2}z/M_*$, we obtain
  \begin{eqnarray}
	z_{\rm dec}\sim \frac{m_{3/2}^2}{M_*}.
 \end{eqnarray}
 Then, assuming the initial amplitude of $z$ to be $z_0\sim M_*$, the
dilution factor is given by $D\sim M_*/m_{3/2}$, and hence
  \begin{eqnarray}
	\frac{n_B}{s} \sim
	D^{-1} \frac{m_{3/2} {\rm Im}(\phi_{0}^{4})}{M_{*}^{2}} 
	\sim
	\left(\frac{m_{3/2}}{M_*}\right)^{1/2}
	\left(\frac{|\phi_0|^4}{M_*^4}\right)
	\sin 4\theta_0.
 \label{nb/s(10TeV)}
 \end{eqnarray}
 Thus, the baryon-to-entropy ratio may be larger than $\sim 10^{-10}$
even after the decay of the Polonyi field.

However the reheating temperature after the decay of $z$ is
likely to be too low. By using Eq.~(\ref{Gamma_z}), the reheating
temperature is estimated as
  \begin{eqnarray}
 	T_{\rm R} \sim g_*^{-1/4}\sqrt{\Gamma_z M_*}
	\sim 1 {\rm MeV} \times 
	\left(\frac{m_{3/2}}{\rm 10 TeV}\right)^{3/2}.
 \end{eqnarray}
 Thus, if the gravitino mass is heavier than about 10 TeV, the Polonyi
field may decay before the big-bang nucleosynthesis (BBN), and the
scenario which gives Eq.(\ref{nb/s(10TeV)}) may be viable.\footnote
 {In fact, even if $m_{3/2}\gtrsim$ 10TeV, there may still be a
problem since the lightest superparticle produced by the decay of the
Polonyi field may overclose the Universe \cite{MYY,KMY}. To solve this
difficulty, we may have to accept a much larger gravitino mass, 
or a scenario
in which the lightest superparticle in the MSSM sector is unstable.}
 However, for a favorable range of the gravitino mass
($m_{3/2}\lesssim$ 1TeV), the reheating temperature is less than 100
keV which is lower than the temperature where the big-bang
nucleosynthesis (BBN) starts.  This means that the decay of $z$
significantly affects the results of the standard BBN scenario.
In this case, we need some mechanism to reduce the energy density of
the Polonyi field. A thermal inflation \cite{LS} is an interesting
candidate for it. The baryon-to-entropy ratio in this case is discussed
in Section~\ref{sec:moduli}. By using the fact that the ratio
of $n_B$ to $\rho_z$ is constant in time, we obtain
 \begin{eqnarray}
	\frac{n_B}{s} \sim \frac{\rho_z}{s}\times
	m_{3/2}^{-1} \left(\frac{|\phi_0|}{M_*}\right)^4 \sin 4\theta_0.
	\label{nb-rhoz}
 \end{eqnarray}
 Thus, once we fix the ratio $\rho_z/s$ after the late inflation, the
baryon-to-entropy ratio can be estimated.

\end{document}